\documentclass[aps,pre,showpacs,preprint,superscriptaddress,nofootinbib]{revtex4-1} 
\usepackage[dvips]{epsfig,graphicx}
\usepackage{graphicx}
\usepackage{float}
\usepackage{setspace}
\usepackage{amsfonts}
\usepackage{amssymb}
\usepackage[mathscr]{euscript}
\usepackage{wasysym}
\usepackage{bm}% bold math
\usepackage{soul}
\RequirePackage{color}

\def\be{\begin{equation}}
\def\ee{\end{equation}}
\def\la{\langle}
\def\ra{\rangle}

\begin{document}

\title[Thermodynamic and topology]{Sensitivity of the thermodynamics of two-dimensional systems towards the topological classes of their surfaces} 
%Thermodynamic and topology:
%specific heat and  magnetic susceptibility 
%of the Ising model are sensitive to topological class of surfaces.}
\author{Oleg A. Vasilyev}\email{present affilation: Emanuel Institute of Biochemical Physics,
Russian Academy of Sciences, Kosygina street 4, 119334 Moscow, Russia }
\affiliation{Max-Planck-Institut f{\"u}r Intelligente Systeme,
 Heisenbergstra{\ss}e~3, D-70569 Stuttgart, Germany}
\affiliation{IV.  Institut f\"ur Theoretische Physik,  Universit{\"a}t Stuttgart,  Pfaffenwaldring 57, D-70569 Stuttgart, Germany}

\author{Anna Macio\l ek }\email{maciolek@is.mpg.de }
\affiliation{Max-Planck-Institut f{\"u}r Intelligente Systeme,
 Heisenbergstra{\ss}e~3, D-70569 Stuttgart, Germany}
\affiliation{Institute of Physical Chemistry, Polish Academy of Sciences, Kasprzaka 44/52, PL-01-224 Warsaw, Poland}
\author{S. Dietrich}
\affiliation{Max-Planck-Institut f{\"u}r Intelligente Systeme,
 Heisenbergstra{\ss}e~3, D-70569 Stuttgart, Germany}
\affiliation{IV.  Institut f\"ur Theoretische Physik,  Universit{\"a}t Stuttgart,  Pfaffenwaldring 57, D-70569 Stuttgart, Germany}

%{Lattice theory and statistics (Ising, Potts, etc.)}
%{Critical point phenomena}
%{Monte Carlo methods}

\today

\begin{abstract}
Using Monte Carlo simulations we study  the two-dimensional Ising model on  triangular,  square, and hexagonal lattices 
with various topologies. We focus on the behavior of the magnetic susceptibility and of the specific heat near the critical point of
the  planar bulk system.
We find that scaling functions of these quantities
on the spherical surface (Euler
characteristic $\mathcal{K}=2$) differ from the scaling functions on
the projective plane ($\mathcal{K}=1$) which, in turn, differ from
the scaling functions on the torus and on the Klein bottle
(both $\mathcal{K}= 0$).
This provides strong evidence  that phase transitions of the Ising model on  two-dimensional surfaces
 depend on their topologies.  
\end{abstract}

\maketitle
\section{Introduction}
\label{sec:introduction}

Whether topology can drive  modifications of phase transitions is  an  intresting issue in  basic physics~\cite{Moshchalkov95}. It is also relevant 
for actual systems because   topological surfaces can form either spontaneously, such as the  membrane of vesicles in biological systems,
or they can be fabricated~\cite{AdvMat,Dobrovolskiy22} like, \textit{e.g.}, 
the M\"obius rings,  from   micro-sized single crystals~\cite{tanda2002crystal} and from self-assembled chiral  block copolymers~\cite{chirial_Moebius,Quyang}.

Experiments show  that lipid membranes  forming spherical  plasma vesicles  undergo a  second-order phase transition which belongs
to the two-dimensional (2D) Ising universality class~\cite{Veatch,GKMV}. A phase transition of the same type could also occur on the M\"obius ring,
if a ferromagnetic material would be used for its fabrication~\cite{Pylypovskyi}. 
This has  motivated us to study  the Ising model on finite
2D manifolds exhibiting  various topologies.  For that purpose we have used  Monte Carlo (MC) simulations, focusing on the vicinity of 
the 2D bulk critical point.  
In a  recent Letter~\cite{VMD}, we have  investigated the finite-size scaling function of the magnetic susceptibility 
and showed that it is the same for  the surface of the torus and  of the Klein bottle (Euler characteristic $\mathcal{K}=0$ for both of them) 
but  differs from  the one for the 
projective plane (Euler characteristic $\mathcal{K}=1$) and for the surface of the sphere (Euler characteristic $\mathcal{K}=2$).
This result renders clear evidence   that the universal properties of the continuous order-disorder phase transition in the Ising model for  2D manifolds 
 do depend on their topology.  
 
The present study provides  a  detailed description  of  our non-standard Monte Carlo (MC) simulations presented in Ref.~\cite{VMD},
and it extends our analysis of the dependence 
on the surface topology  to another thermodynamic quantity, {\it i.e.},  the specific heat.
Moreover, in addition to the  triangular and square lattices, which we have used in our aforementioned Letter to form 2D manifolds~\cite{VMD},
here we employ  also hexagonal facets. 

Our systematic study   broadens the knowledge about the dependence of various physical quantities on the topology of the surface.
The understanding accumulated up to here is rather limited.  There are  numerical results for the Binder cumulant and for  the critical magnetization distribution
 on the torus, the M\"obius strip, and the Klein bottle~\cite{KO}. For topologies such as the M\"obius strip and the Klein bottle, also analytical expressions for the partition function are available
~\cite{LW}. 
Earlier studies led to numerical results for various cumulants, the maximum of the specific heat, and the shift 
of the pseudo-critical point  on the torus and on  surfaces which
are topologically equal to a sphere~\cite{HL}.  All these studies reveal distinct results for various topologies.
In particular the finite-size effects turned out to be strongly influenced by the topology.

The presentation is organized as follows. In Sec.~\ref{sec:asp}  we briefly introduce the Ising model in its current nomenclature. 
In Sec.~~\ref{sec:geom}   triangular,  square, and hexagonal decorations of the lattice are described. 
The methods of constructing manifolds with the topology of 
 the Klein bottle, the real projective plane,  and the  spherical geometry are described in the same section. 
In Sec.~\ref{sec:results}  we compare the results for the magnetic susceptibility and for the  specific heat scaling functions
 for various topologies. The last section contains our conclusions.

 In the appendices  we describe   auxiliary results: in
 Appendix~\ref{appendix:aspect}, the  dependence of the magnetic susceptibility on the aspect ratio of the lattice; 
and in  Appendix~\ref{appendix:B},
the  scaling of the specific heat for the square lattice.

\section{Models}
\label{sec:asp}
We consider  an arbitrary graph 
consisting of $N$ vertices (sites) connected by
edges (bonds). 
Each site $j$ ($j=1,2,\dots,N$) of the graph is occupied by an Ising spin $s_{j}=\pm 1$.
The energy of the model is given by 
\be
\label{eq:H}
U = - J \sum \limits_{\{ i,j \}} s_{i}s_{j},
\ee 
where the sum $\{ i, j\}$ is taken over all pairs of nearest neighbor spins 
connected by bonds  in accordance with the graph geometry.
We absorb the interaction constant  $J$ into the dimensionless inverse temperature $\beta=J/(k_{\mathrm B}T)$.

Within  Monte Carlo simulations~\cite{LB}  the 
specific heat $C$ can be  computed as
\be
\label{eq:c_def}
C (\beta)=\beta^2 \left( \la U^{2} \ra-\la  U \ra^{2} \right)/N,
\ee
where $N$ is the total number of spins and  $ \la U^{k} \ra$ 
is  the thermal average of the $k$-th power of the energy.
The angle brackets $\la  \dots \ra$ represent the thermal average   of a quantity $A$:
$\la  A \ra = (1/Z) \sum_{\{s \}}A(s) \exp[-\beta U(s) ] $,
where $Z=\sum_{\{s \}} \exp[-\beta U(s) ]$ is the partition function
and the  sum $\sum_{\{s \}}$ is taken over all spin configurations $s=(s_1,\cdots, s_N)$.
We also compute the magnetic susceptibility 
$\chi$ of a system: 
\be
\label{eq:chi_def}
\chi (\beta)=\beta \left( \la M^{2} \ra-\la  M \ra^{2} \right)/N,  
\ee
where  $ \la M^{k} \ra =  \left< \left|\sum_{j} s_{j} \right|^{k} \right>$, $k=1,2$ 
 is the thermal average of the $k$-th
 power of the magnetization.

Our goal is to study the thermodynamic properties of the  Ising model 
on two-dimensional lattices which exhibit distinct topologies. Some of these  topologies can 
be obtained from a square or a (rectangularly) shaped lattice by ``glueing''  opposite edges such that their orientations are parallel.
We plot the Klein bottle, the real projective plane, and the torus
in the first row of Figs.~\ref{fig:shape}(a),~(b), and (c), respectively.
 The directions, in which opposite sides of a rectangular lattice
 should be glued  in order to produce  such topologies, are presented in the second row of this figure.
In turn, the  polygon subjected to the glueing  (of square or rectangular shape)
may have different decorations (\textit{i.e.}, site positions and bond arrangements).
In Fig.~\ref{fig:geom_plane} we plot examples of triangular, square, and 
hexagonal decorations.
We distinguish between the lattice type (triangular, square, hexagonal) and the lattice shape (rectangular or square). 
On the triangular lattice (tri., $\triangle$) each spin has six bonds, 
on the square lattice (sq., $\square$) each spin has four bonds, and 
on the hexagonal  lattice (hex., $\hexagon$ ) each spin has three bonds.

As a reference  case we consider the Ising model defined on a two-dimensional rectangular lattice 
of the size $L_{1} \times L_{2}$ with periodic boundary conditions (\textit{i.e.}, a torus) having  a certain bond arrangement 
(triangle, square, and hexagonal) with  lattice spacing $a=1$.
We measure the size of a lattice  in terms of the   number $L_1$    of spins  in a row (horizontal layers)
and by the number $L_2$ of vertical layers. 
We note that  a square lattice of  rectangular shape $L_1 \times L_2$ has the geometrical size  $aL_{1} \times aL_{2}$, whereas for  a triangular lattice of size $N = L_1 \times L_2$
the geometrical size is $aL_1 \times \frac{\sqrt{3}}{2}aL_2$.

In accordance with  finite-size scaling theory~\cite{Barber,Privman}, for  the square 
lattice with the {\it square} shape $L_{1}=L_{2}=L$ and  with  the total number  $N=L^{2}$ of spins,   the magnetic susceptibility can  be expressed as 
$
\chi(\beta,L)= L^{\gamma/\nu}  {\mathcal X}(tL^{1/\nu})= 
N^{\gamma/(2\nu)} {\mathcal X}(t N^{1/(2\nu)}).
$
$\mathcal X$ is the dimensionless magnetic susceptibility scaling function,
  $t=(T-T_c)/T_c=(\beta_c-\beta)/\beta$  
  is the reduced temperature,  and $\nu=1$ and $\gamma=7/4$ are the correlation 
  length and the magnetic susceptibility critical exponents~\cite{PV}, respectively.

For  the square lattice 
of {\it rectangular} shape with the aspect ratio $\rho=L_{1}/L_{2} \ne 1$, or with
 $L_{1}=\rho L_{2}=:\rho L $, the magnetic susceptibility 
can be expressed as
\be
\label{eq:chi_scal_sq}
\chi(\beta,\rho L,L)=N^{\gamma/(2\nu)}
 {\mathcal X}(\tau ,\rho),
\ee
Here,  $N =L_{1}L_{2}= \rho L^{2}$ is the total number of spins and 
$\tau=t N^{1/(2\nu)}$ is the temperature scaling variable. 

We shall also consider the regular triangular lattice  and the 
regular hexagonal  lattice  with the aspect ratio $\rho=L_{1}/L_{2}$ 
close to  the special value $r_{0}$. This value $r_{0}$ is equal to $\sqrt{3}/2$ for the regular triangular lattice and $\sqrt{3}$ for   the  regular hexagonal  lattice.
The reason for  selecting these values of the aspect ratio  is discussed in Appendix~\ref{appendix:aspect}.
One finds that  lattices with these aspect ratios 
(in terms of the number of spins) exhibit a square geometrical shape.

We  shall use the  following values for the critical amplitudes of the bulk correlation length  as obtained from the second moment of the two-point correlation function 
in the region above the critical point: $\xi_0^+\simeq 0.525315$ for the triangular ($\triangle$)
lattice, $\xi_0^+\simeq 0.56706$
 for the square ($\square$) lattice,
and  $\xi_0^+\simeq 0.657331$ for the hexagonal ($\hexagon$) lattice~\cite{BCG}.

\section{Lattices and topologies}
\label{sec:geom}

Topology of a  polyhedra is classified in accordance with its Euler characteristic~\cite{Flegg}
\be
\label{eq:euler}
{\mathcal  K}=N-E+F, 
\ee
where $N$ is the number of vertices (spins),
$E $ is the number of edges (bonds), and $F $ is the number of facets. Every  convex polyhedron can be turned into a connected, simple, planar graph. In turn, each
connected planar graph  has the same Euler characteristic ${\mathcal K}=2$.

\begin{figure*}[h]
\centerline{\includegraphics[width=0.8\textwidth]{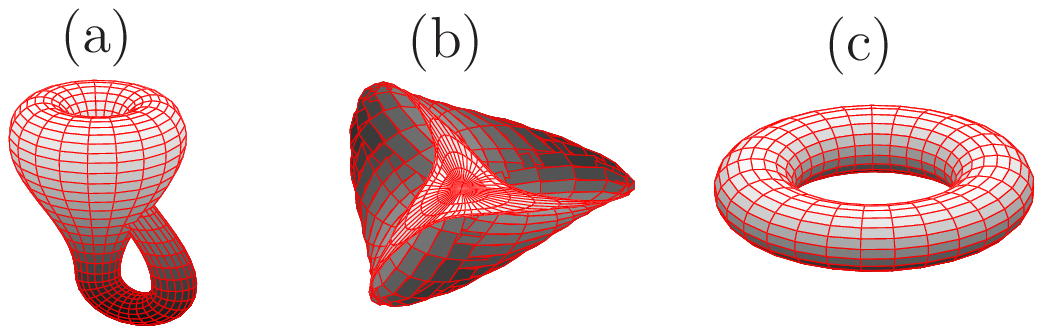}}
\centerline{\includegraphics[width=0.8\textwidth]{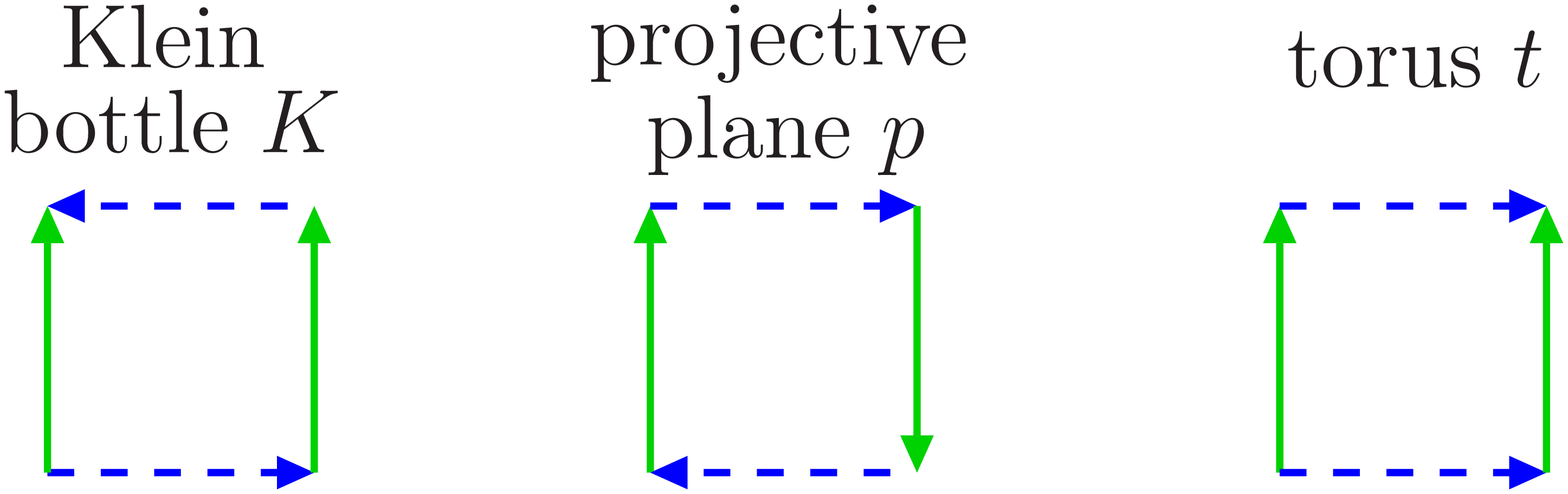}}
\caption{
Top row: two-dimensional representations of (a) the Klein bottle $K$;
of (b) the real projective plane $p$ (known also as the Roman surface or Steiner surface~\cite{Steiner});
of (c) the torus $t$, all
immersed in the three-dimensional $(x,y,z)$ space (see Fig.~2).
The gray-scale code for surface facets denotes the position
   of the facet in z direction. Red lines denote the positions of the constant parameters for the surface parameterization.
In order to produce these surfaces, the corresponding blue and green edges of
 a  square should be ``glued'' together with the arrows matching (bottom row).
Dashed blue lines and solid green lines represent the opposing boundaries. 
}
\label{fig:shape}
\end{figure*}

\begin{figure*}[h]
\centerline{
\mbox{
\includegraphics[width=0.35\textwidth]{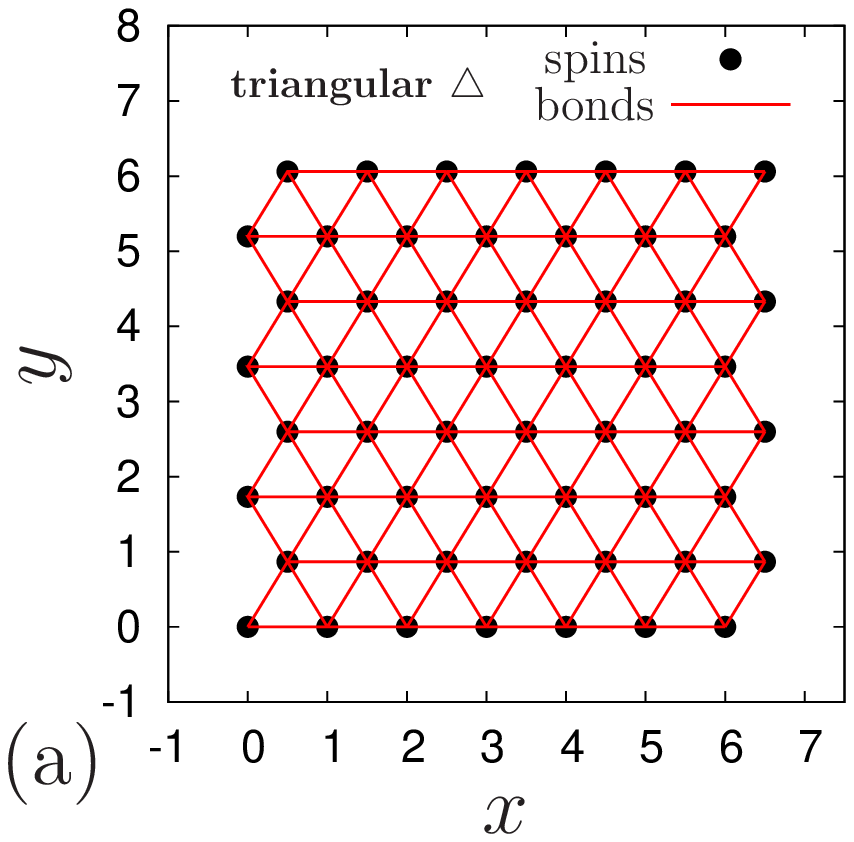}\hspace{-0.01\textwidth}
\includegraphics[width=0.35\textwidth]{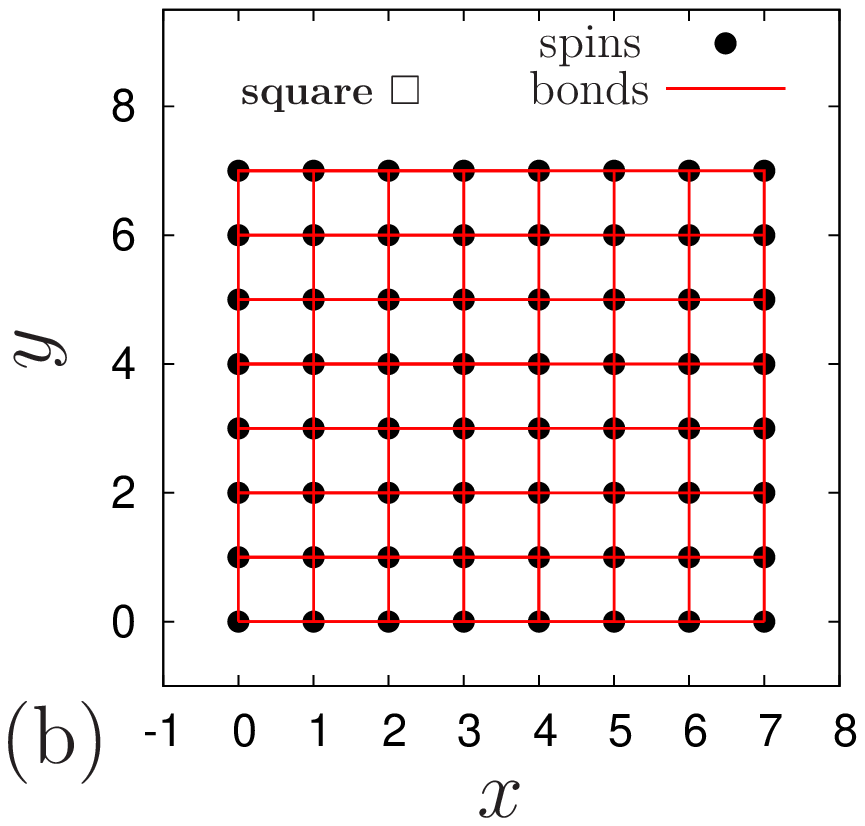}\hspace{-0.01\textwidth}}}
\centerline{
\mbox{
\includegraphics[width=0.35\textwidth]{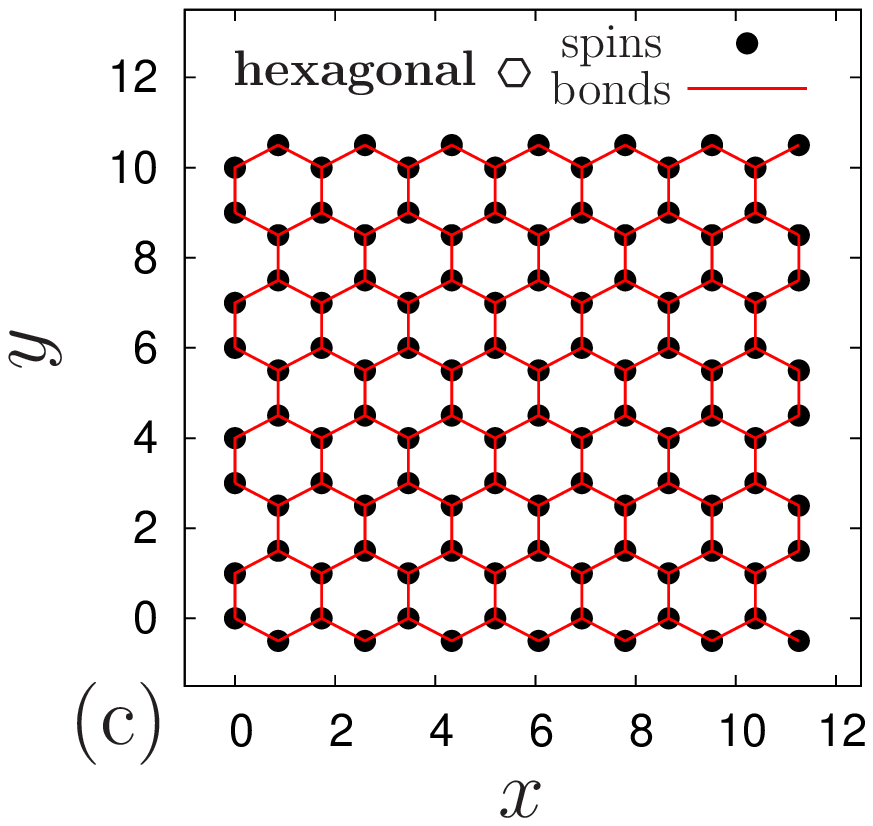}
\includegraphics[width=0.35\textwidth]{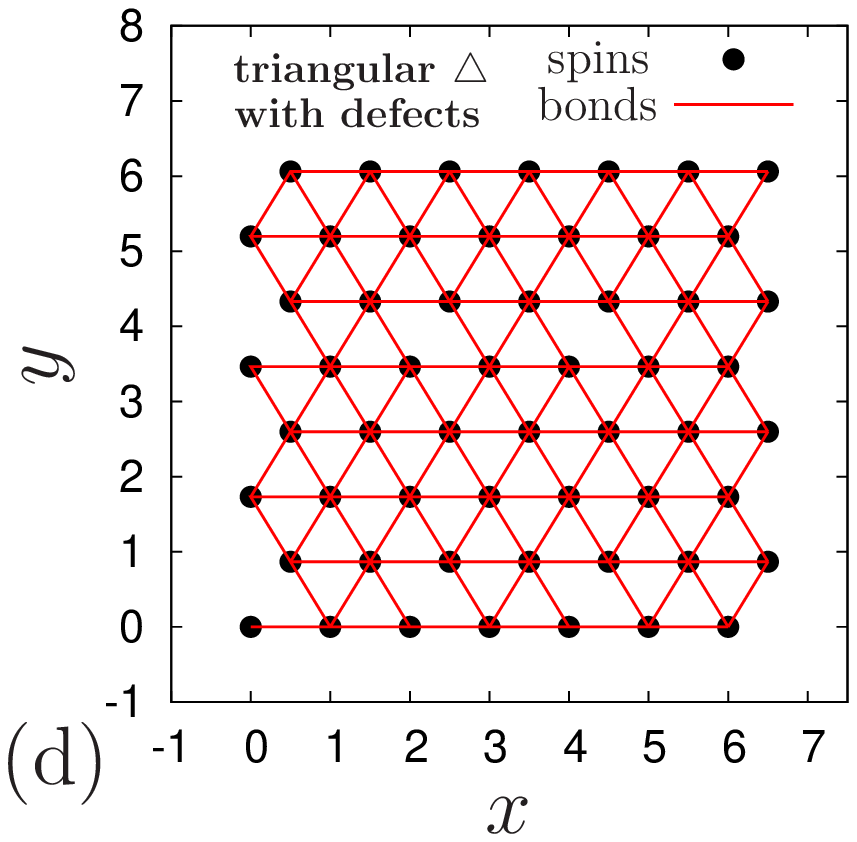}}
}
\caption{ Decorations of  lattices:
(a)  regular triangular lattice ($\triangle$) of size $7 \times 8$, {\it i.e.},    $N=56$;
(b)  regular square lattice ($\square$)  of size $8 \times 8$, {\it i.e.}, $N=64$;
(c)  regular hexagonal lattice ($\hexagon$)  of size $14 \times 8$, {\it i.e.}, $N=112$;
(d)  triangular lattice ($\triangle$) with 12 defects of type $b=5$ ({\it i.e.}, 6 missing bonds) 
of size $7 \times 8$, {\it i.e.},    $N=56$.
}
\label{fig:geom_plane}
\end{figure*}

For a regular triangular lattice  forming  a torus (which cannot be turned into a connected planar graph), 
exactly six edges (bonds, $b=6$) meet at each vertex (here and later on we denote by the letter $b$ the number of bonds (edges). 
Each edge has two ends so that
the total number of edges is $E=3N$. Each triangular facet 
contains 3 vertices  and  each vertex participates  in 6 facets,
 so that the total number of facets is $F=6N/3=2N$.
 Therefore, the Euler characteristic of the torus   is
${\mathcal  K}_{\mathrm{torus}}=0$.
We can form the  torus by using the triangular  decoration (see Fig.~\ref{fig:geom_plane}(a))
 and by applying the appropriate boundary conditions from Fig.~\ref{fig:shape}(c).
The Klein bottle (with Euler characteristic ${\mathcal  K}_{\mathrm{Klein}}=\mathcal{K}_{\mathrm{torus}}= 0$) is  constructed 
 by glueing the edges along the directions shown in   Fig.~\ref{fig:shape}(a).
The real projective plane (${\mathcal  K}_{\mathrm{p.\;plane}}=1$)
may be produced by glueing edges in opposite directions for both pairs of sides (see Fig.~\ref{fig:shape}(b)).
For the Klein bottle and the real projective plane 
Eq.~(\ref{eq:euler}) is not applicable,  because   these 
non-orientable surfaces cannot be realized  in three-dimensional space without self-intersection. 

There is an analogous  procedure for the square lattice: 
we use the square decoration of the lattice (see Fig.~\ref{fig:geom_plane}(b)) and apply  the appropriate  glueing of opposite sides (see Fig.~\ref{fig:shape}).
 In the case of the projective plane 
two pairs of sites in opposite corners form double bonds between them.
The same procedure is valid for the hexagonal lattice: one glues 
opposite sides of the lattice with hexagonal facets (see Fig.~\ref{fig:geom_plane}(c))
in accordance with one of the orientations indicated in Fig.~\ref{fig:shape}.

\begin{figure*}[h]
\centerline{
\mbox{
\includegraphics[width=0.31\textwidth]{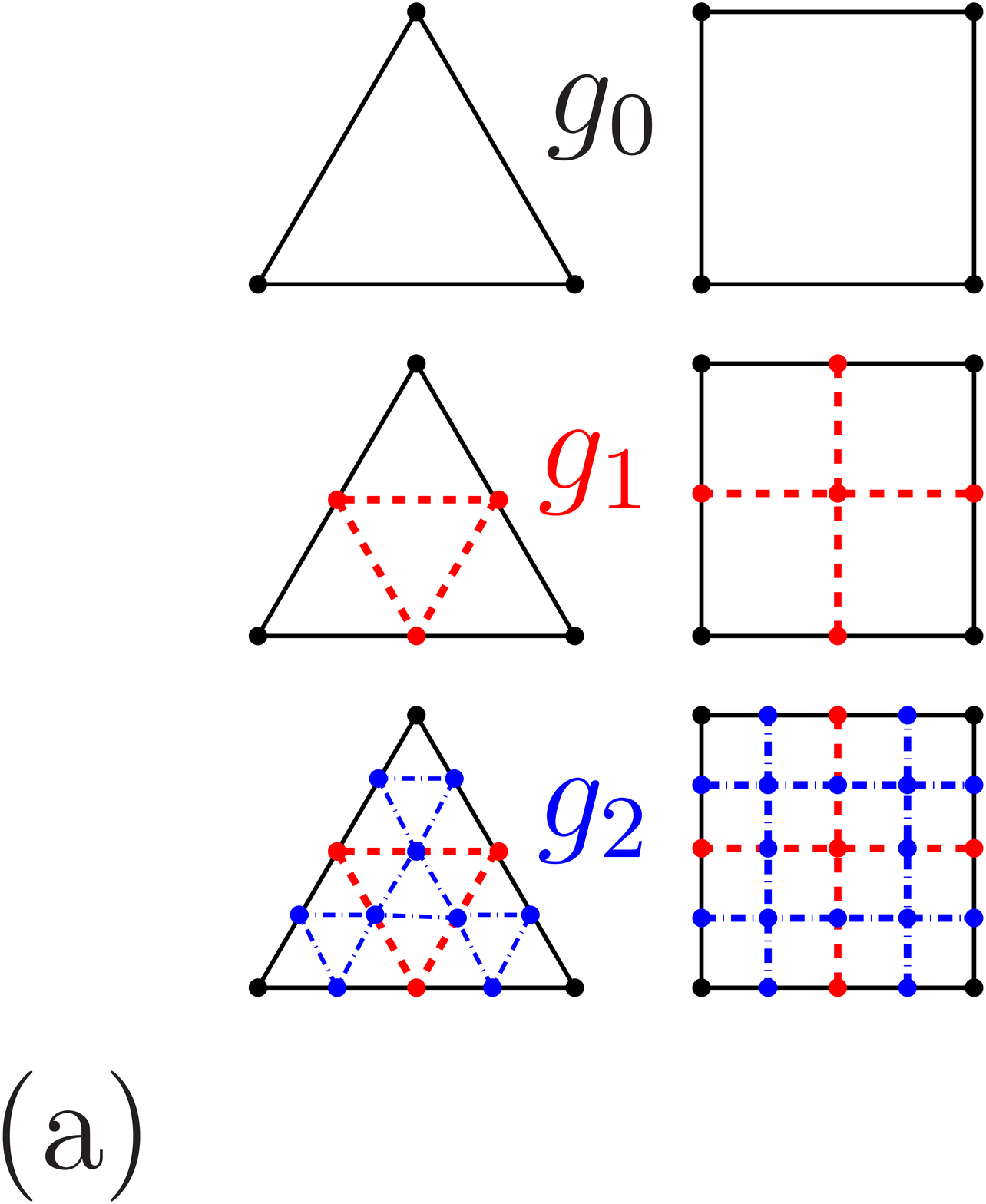}
\includegraphics[width=0.31\textwidth]{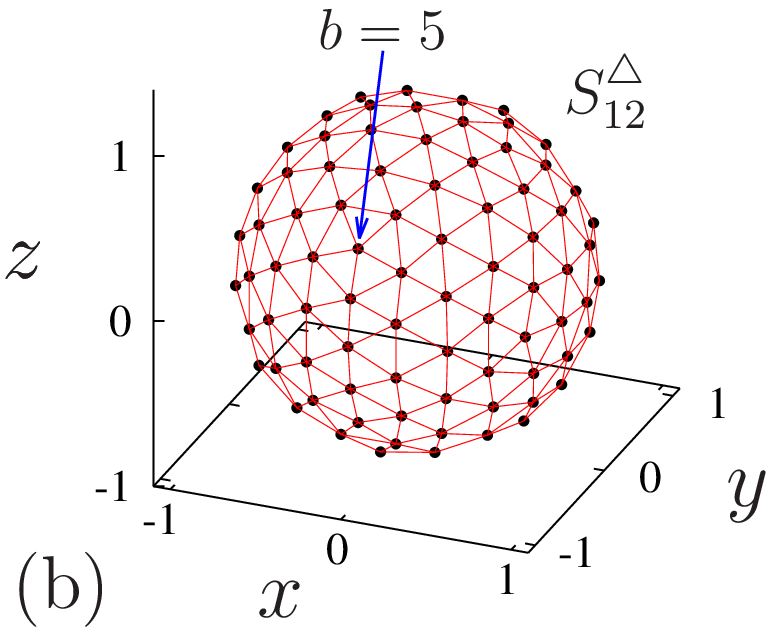}
\includegraphics[width=0.31\textwidth]{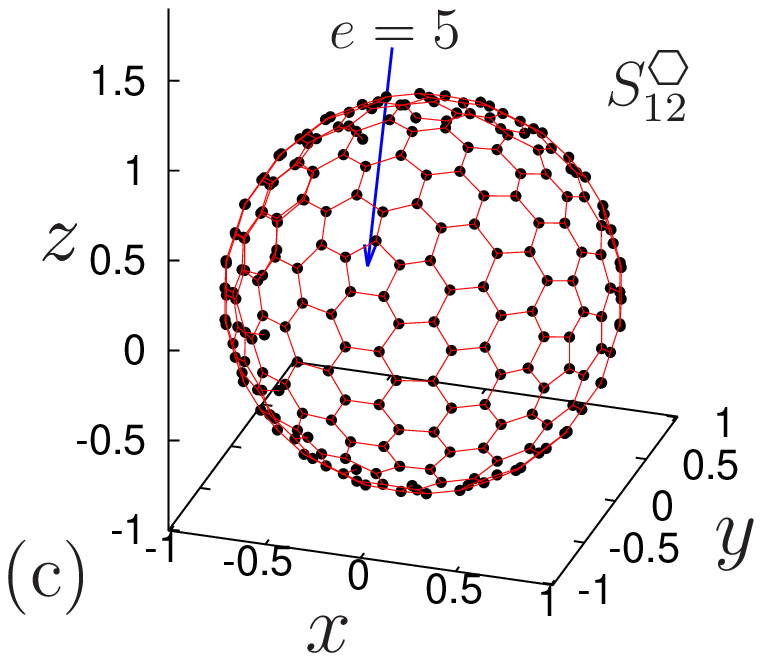}
}}
\centerline{
\mbox{
\includegraphics[width=0.31\textwidth]{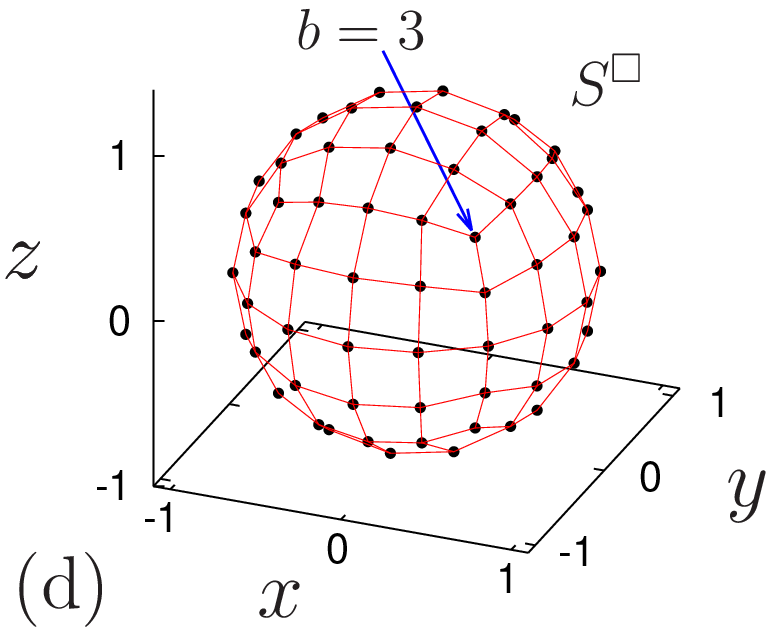}
\includegraphics[width=0.31\textwidth]{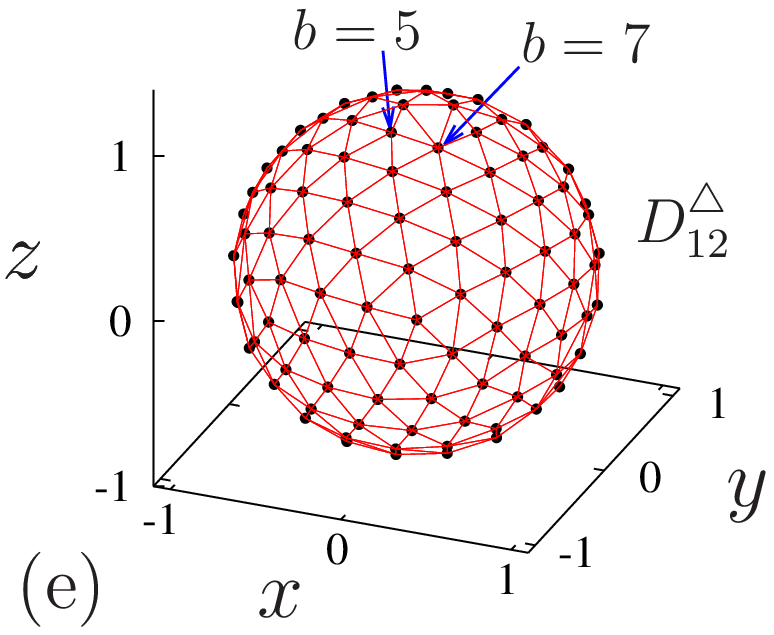}
\includegraphics[width=0.31\textwidth]{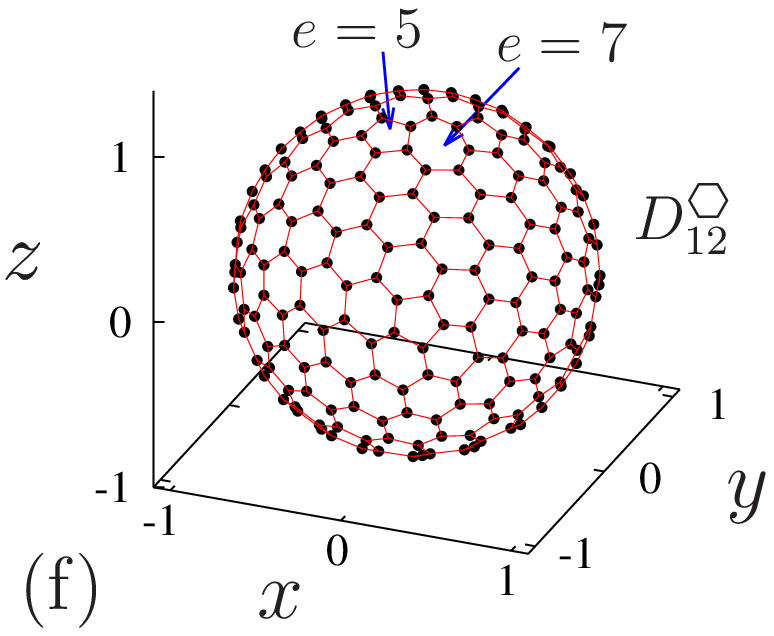}
}
}
\caption{Structure of lattices with spherical geometry  ($S$):
(a) Recursive triangulation of a single triangular facet: the initial facet for the zeroth generation 
($g_0$, solid line), the first ($g_1$, dashed line), and the second
($g_2$, dash-dotted line) generation.
(b) Regular triangulation of the sphere ($S^{\triangle}_{12}$) with $n_{5}=12$
defects of type $b=5$ and $N=162$ (one defect $b=5$ is indicated by the blue arrow).
(c) Regular decoration  of the sphere by hexagons
($S^{\hexagon}_{12}$)    with $n_{5}=12$
defects of type $e=5$, where $e$ denotes the number of \textit{e}dges, and $N=320$ (one defect $e=5$ is indicated).
(d) Spherical representation of the surface of a cube with a regular 
square lattice  ($S^{\square}$, see Table II)  with $n_3=8$ defects of type $b=3$ (corners) and
 $N=98$ (one defect $b=3$ is indicated).
(e) Random triangulation of the sphere ($D^{\triangle}_{12}$) with 
 $n_{5}=16$ defects of type $b=5$ and $n_{7}=4$ defects of type $b=7$ and
 $N=162$ (two defects $b=7$ and $b=5$ are indicated).
(f) Random representation of the sphere by hexagons
($D^{\hexagon}_{12}$, dual to $D^{\triangle}_{12}$) with 
 $n_{5}=16$ defects of type $e=5$ and $n_{7}=4$ defects of type $e=7$ and 
 $N=320$. 
The new bond on the dual lattice is perpendicular to the corresponding bond
   on the original lattice, \textit{i.e.}, a vertex is transformed into a facet and vice versa.
For example, the defect vertex with $b=5$ bonds is transformed into the defect
   pentagon facet with $e=5$ edges (two defects $e=7$ and $e=5$, which correspond to panel (e), are indicated). 
  }
\label{fig:geom_sph}
\end{figure*}

Now we turn to  surfaces with spherical topology corresponding to  the Euler characteristic 
${\mathcal  K}_{\mathrm{sphere}}=2$. 
If we  denote $n_b$ as the number of vertices which give rise to exactly $b$ edges,  one has for the  total number of vertices  $N =\sum \limits_{b}n_b$.
Since  the total number of edges  is $E= \frac{1}{2}\sum \limits_{b} b n_b $  and 
the total number of triangular facets is $F=  \frac{1}{3} \sum \limits_{b} b n_{b}$, 
we can rewrite Eq.~(\ref{eq:euler}) as 
\begin{equation}
\label{eq:euler1}
 {\mathcal  K}=N-E+F=\sum \limits_{b}(1-b/6) n_b.
\end{equation}
From this expression it  follows that for spherical surfaces the regular triangulation, i.e.,  by a lattice of 
 vertices which participate in six-bonds (\textit{i.e.}, $b=6$) vertices,  as  used for the surface of a torus, is not possible.
For a sphere one  needs  a fraction of vertices with  the associated number of bonds less than $6$,  
in accordance with  Eq.~(\ref{eq:euler1}) and with  $\mathcal{K}=2$:
$\sum \limits_{b}(6-b) n_b=12$;
for example, $n_{5}=12$ vertices with $b=5$
 bonds, $n_{4}=6$ vertices with $b=4$ bonds etc.
We can construct the regular triangulation of a sphere
in the following way. We  start with a regular\footnote{One definition of a regular polyhedron is, that the faces are congruent regular polygons which are assembled in the same way around each vertex.} seed polyhedron
with triangular facets (tetrahedron, octahedron, or icosahedron) 
and  a circumscribed sphere of unit radius as a zero generation 
$k=0$ of our triangulation. In the next step,  we add 
new vertices in the middle of each edge and connect these new vertices by  
new edges  in order to create  a generation  $g_1$ (see Fig.~\ref{fig:geom_sph}(a)). The initial facet for the zero generation $g_0$
is a triangle. After this starting procedure,
each triangle facet is split into four new facets 
of triangular shape $g_1$. Each new vertex is linked to exactly $b=6$ edges. 
After each iteration step we project the coordinates 
of the   vertices of the  next generation onto the surface of a unit sphere. After that we continue with the second
iteration $g_2$, and so on.
The number of vertices $N(k)$ for the $k$-th iteration step depends on the  type of the polyhedron.  
We repeat this procedure until the desired level of triangulation is achieved.
Because this recursive procedure generates only  vertices with $b=6$,
each  triangulation   inherits the number of defects from the seed polyhedron.
The triangulation, which starts from a \textit{tet}rahedron ( denoted as  $S^{\triangle}_4$; here and in the following  the subscript indicates the number of vertices of the seed polyhedron, and 
$S$ stands for sphere),  has $n_{3}=4$ defects
with $b=3$ bonds and has $N_{\mathrm{tet.}}(k)=4+2(4^k-1)$ vertices in the $k$-th step of the iteration.
The triangulation, which starts from an \textit{oct}ahedron  ( denoted as  $S^{\triangle}_6$) 
has $n_{4}=6$ defects with $b=4$ bonds and 
has $N_{\mathrm{oct.}}(k)=6+4(4^k-1)$ vertices in the  $k$-th step of the iteration.
Another regular triangulation,  which  starts from  an \textit{ico}sahedron  (denoted as $S^{\triangle}_{12}$) 
has $n_{5}=12$ defects with $b=5$ bonds and has $N_{\mathrm{icos.}}(k)=12+10(4^k-1)$ 
vertices in the $k$-th step. The latter version of triangulation is expected
 to be more homogeneous due to the
 small degree  $(6-b)$ of defects, which leads to a higher level of symmetry. In Fig.~\ref{fig:geom_sph}(b) we plot the generation
$k=2$  of the $S^{\triangle}_{12}$ triangulation, which contains $N=162$
vertices. One of the $b=5$ defects is indicated by an arrow.
\begin{table*}
\caption{Structure of regular decorations of a sphere: number of vertices $N(k)$,
edges $E(k)$, and facets $F(k)$ for the $k$-th generation of triangulations $\triangle$, which started from 
a tetrahedron, an octahedron, and an icosahedron. The square decoration $\square$
starts from a cube, and  the hexagonal decoration $\hexagon$ is  obtained 
as being dual to the triangular lattice for the icosahedron.
}
\label{tab:NEF}
\begin{center}
\begin{tabular}{|c|c|c|c|c|c|}
\hline
\hline
 seed  polyhedron &  facet  & gen.   &  $N(k)$ & $E(k)$   & $F(k)$ \\
\hline
tetrahedron& $\triangle$ &   k  &  $ 4 +2 ( 4^{k}-1)$      &   $ 6 \times 4^{k}$    &  $ 4 \times 4^{k}$   \\
\hline
octahedron& $\triangle$ &   k  &  $ 6 +4 ( 4^{k}-1)$      &   $ 12 \times 4^{k}$    &  $ 8 \times 4^{k}$   \\
\hline
icosahedron& $\triangle$ &   k  &  $ 12 +10 ( 4^{k}-1)$      &   $ 30 \times 4^{k}$    &  $ 20 \times 4^{k}$   \\
\hline
cube& $\square$ &   k  &  $ 2 + 6 \times 4^{k}$      &   $ 12 \times 4^{k}$    &  $ 6 \times 4^{k}$   \\
\hline
dual to icosahedron& $\hexagon$ &   k  &   $ 20 \times 4^{k}$     &   $ 30 \times 4^{k}$    & $ 12 +10 ( 4^{k}-1)$     \\
\hline
\hline
\end{tabular}
\end{center}
\end{table*}

The regular decoration of a sphere by hexagons $S^{\hexagon}_{12}$
is constructed as a dual lattice for the triangular decoration,  \textit{i.e.}, 
in the dual hexagonal lattice, each vertex  corresponds to the triangular facet 
of the seed lattice and has exactly three edges.
Defects of the seed lattice are transformed into facets, which are not hexagons.
In  Fig.~\ref{fig:geom_sph}(c) we plot the decoration by hexagons 
which is dual to the triangular decoration shown in Fig.~\ref{fig:geom_sph}(b).
The defect with five \textit{e}dges ($b=5$) is transformed into a defect pentagonal facet denoted by $e$
with  $e=5$. 

For the square lattice we repeat the same procedure, using the cube as a seed polyhedron. 
In the first step, every square facet  $g_0$ of a cube is split into four squares
 $g_1$ by adding new vertices and edges (see Fig.~\ref{fig:geom_sph}(a)).
In the next step, each of  the previously generated $g_1$ facets is split into  four
new facets $g_2$ and so on. After each iteration step we project  the new vertices onto the unit sphere.
This way we obtain a ``sphero-cube''  $S^{\square}$, \textit{i.e.}, a surface with spherical topology
decorated by square facets as done in Ref.~\cite{HL}. 
This object has $n_{3}=4$ defects with $b=3$ bonds.
The number of vertices  in the  $k$-th step is $N_{\mathrm{cube}}(k)=2+6\times 4^k$. 
In Fig.~\ref{fig:geom_sph}(d) we plot the ``sphero-cube'' with $N=98$ for the generation $k=2$;
one of the $b=3$ defects is indicated by an arrow. 
%From this plot it is evident that it is not possible to cover the sphere by regular square lattice.
 The structure of regular decorations of a sphere is summarized in Table~\ref{tab:NEF}.
\begin{table*}
\caption{List of the studied  geometries for 
triangular (tri., $\triangle$),  square (sq., $\square$), and hexagonal (hex., $\hexagon$)
decorated lattices. Symbols for the notations,  Euler
characteristic $\mathcal K$, geometries, and brief descriptions (boundary conditions bc; random triangulation of the sphere $D^{\triangle}_{12}$.)}
\label{tab:geom}
\begin{tabular}{|c|c|c|l|l|}
\hline
type  & sym. &   $\mathcal K$ & geometry & description  \\
\hline
\hline
tri. & $T^{\triangle}$   & 0 & torus   &  triangular lattice with toroidal bc              \\
\hline
tri. & $T^{\triangle}_{d}$   & 0 & torus   &  triangular lattice with defects,   toroidal bc           \\
\hline
tri. & $T^{\triangle}_{r}$   & 0 & torus   &  triangular lattice with defects, toroidal bc              \\
\hline
tri.  &$K^{\triangle}$   & 0 & Klein bottle   &  triangular lattice with Klein bottle bc       \\
\hline
tri.  &$P^{\triangle}$   & 1 &  projective plane   & triangular lattice with real projective plane bc       \\
\hline
tri.  &$S^{\triangle}_{12}$  & 2  & sphere    & triangulation of the sphere with  $n_{5}=12$ defects              \\
\hline
tri.  &$S^{\triangle}_{6}$  & 2  & sphere    & triangulation of the sphere with  $n_{4}=6$ defects              \\
\hline
tri.  &$S^{\triangle}_{4}$  & 2  & sphere    & triangulation of the sphere with  $n_{3}=4$ defects              \\
\hline
tri.  &$D^{\triangle}_{12}$ & 2    & sphere   & dynamic triangulation of the sphere  \\
\hline
\hline
sq. & $T^{\square}$   & 0 & torus   &  square lattice with toroidal bc              \\
\hline
sq.  &$K^{\square}$   & 0 & Klein bottle   &  square lattice with Klein bottle bc       \\
\hline
sq.  &$P^{\square}$   & 1 &  projective plane   & square lattice with  projective plane bc       \\
\hline
sq.  &$S^{\square}$   & 2 & sphere   &  square lattice on the surface of the cube  \\
\hline
\hline
hex. & $T^{\hexagon}$   & 0 & torus   &  hexagonal lattice with toroidal bc              \\
\hline
hex.  &$K^{\hexagon}$   & 0 & Klein bottle   &  hexagonal lattice with Klein bottle bc       \\
\hline
hex.  &$P^{\hexagon}$   & 1 &  projective plane   & hexagonal lattice with real projective plane bc       \\
\hline
hex.  &$S^{\hexagon}_{12}$  & 2  & sphere    &  hexagonal decoration of a sphere (dual to $S^{\triangle}_{12}$)              \\
\hline
hex.  &$D^{\hexagon}_{12}$ & 2    & sphere   & hexagonal decoration of a sphere  (dual to $D^{\triangle}_{12}$) \\
\hline
\hline
\end{tabular}
\end{table*}

From a very simplified point of view the sphere triangulation $S^{\triangle}_{12}$ 
is the lattice   in which 12 vertices have $b=5$
bonds and all remaining  vertices have $b=6$ bonds. 
It is interesting to see whether differences in  the results obtained for 
triangular toroidal lattices (defect-free)  and for spherical lattices (with defects) are due to these defects.
 In order to clarify this  point we  consider a toroidal triangular 
lattice with 6 removed bonds (defects). The presence of those defects
produces 12 vertices with $b=5$ bonds (like for the sphere triangulation $S^{\triangle}_{12}$).
We consider the case,  in which these defect bonds are regularly distributed in order
to have  maximal separation between them  (we denote this case as $T^{\triangle}_{d}$) (Fig.~\ref{fig:geom_plane}(d)).
The case, in which these defect bonds are 
randomly distributed over the lattice, is  denoted as $T^{\triangle}_{r}$.

We have also studied a random (dynamic) triangulation of the sphere. 
In order to create a configuration with random positions of the vertices, we implement Langevin dynamics for $N$ particles  on  a sphere with mutual pairwise repulsion.
Having the vertices randomly distributed,  we apply   Delaunay triangulation~\cite{Delaunay,DelProg}.  (This procedure is described in full details in Ref.~\cite{VMD};
therefore we refrain to repeat it here.) 
%as described in detail  in Appendix~\ref{appendix:dyn}.
The number of particles ( \textit{i.e.}, vertices)  is chosen as for the regular 
triangulation $S^{\triangle}_{12}$.  Accordingly, we denote the  results
for this  \textit{d}ynamic triangulation as $D^{\triangle}_{12}$.
An example of the random triangulation of the sphere with $N=162$
vertices (spins) is shown in Fig.~\ref{fig:geom_sph}(e). This realization 
encompasses  $n_{5}=16$ defects of type $b=5$ and $n_{7}=4$ defects of type $b=7$.
Two of these defects ($b=7$ and $b=5$) are indicated in the figure.
An example of  hexagonal decoration (dual to the lattice in Fig.~\ref{fig:geom_sph}(e))
is provided in Fig.~\ref{fig:geom_sph}(f). Here pentagon  $e=5$ and heptagon $e=7$
facets  correspond to $b=5$ and $b=7$ defects in Fig.~\ref{fig:geom_sph}(e).
Notations and brief descriptions for the considered geometries are
collected in Table~\ref{tab:geom}.

\section{Simulation results}
\label{sec:results}

We use Monte Carlo (MC) simulations~\cite{LB} to compute the specific heat $C$ and the magnetic susceptibility $\chi$
in accordance with Eqs.~(\ref{eq:c_def}) and (\ref{eq:chi_def}), respectively,  for various lattices 
with distinct topologies.
Our aim is to determine the finite size scaling functions of these quantities
near  the bulk  critical point of the two-dimensional Ising model.
The specific heat $C$ and the magnetic susceptibility $\chi$ are  second derivatives of the free energy,  and at the critical point  they 
diverge as the system size (i.e., the number of spins $N$) tends to infinity $N \to \infty $, \textit{i.e.},  $C \propto \ln(N)$ and  $\chi \propto N^{7/8}$.
(The critical scaling of the specific heat is discussed in Appendix \ref{appendix:B}).  
 Quantities like the mean magnetization $\left< | M | \right>/N$
and the mean energy $\la U \ra/N$ per site, which can be easily inferred from the MC simulations,   are not very informative, because they  remain finite
at the critical point. 

Before starting the numerical  measurements, we allow 
the system to thermalize,  using $5 \times 10^6$  (MC) steps. 
Then we perform the thermal average of the thermodynamic quantities
over $10^7$ (MC) steps, split into 10 series
of $10^6$ steps in order to determine the numerical error.
Each MC step consists of a cluster update 
in accordance with the Wolff  algorithm~\cite{Wolff}.
We mainly focus on  the triangular lattice decoration,
because triangulation is a common method for 
the grid representation of surfaces.
We use the temperature scaling variable  $x=\tau/(\xi_0^+)^{1/\nu}= t (N^{1/2}/\xi_0^+)^{1/\nu}$,  where the
numerical values of the bulk correlation length amplitudes $\xi_0^+$ are provided in Sec.~\ref{sec:asp}.
We note that in the special case of a square decoration of the lattice, which is shaped as a square $N=L\times L$,
this scaling variable equals $x=(L/\xi)^{1/\nu}$.

\begin{figure*}[h]
\includegraphics[width=0.49\textwidth]{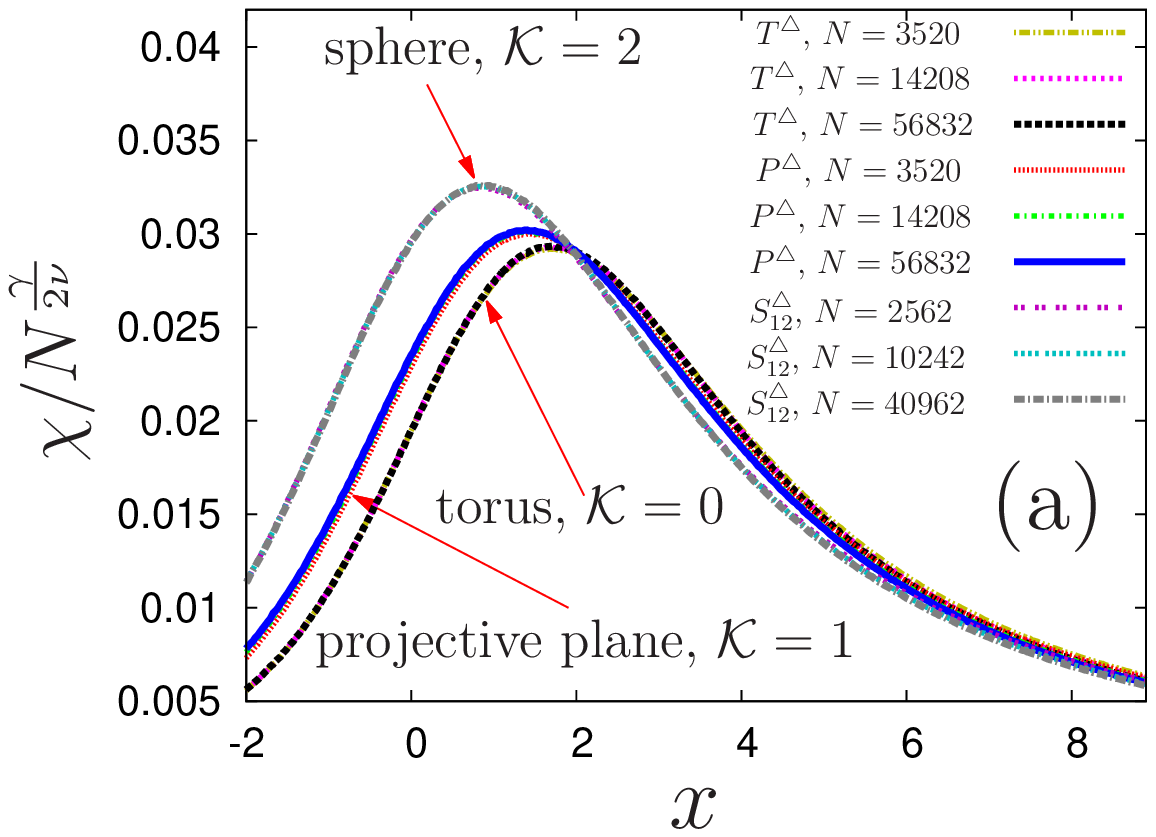}
\includegraphics[width=0.49\textwidth]{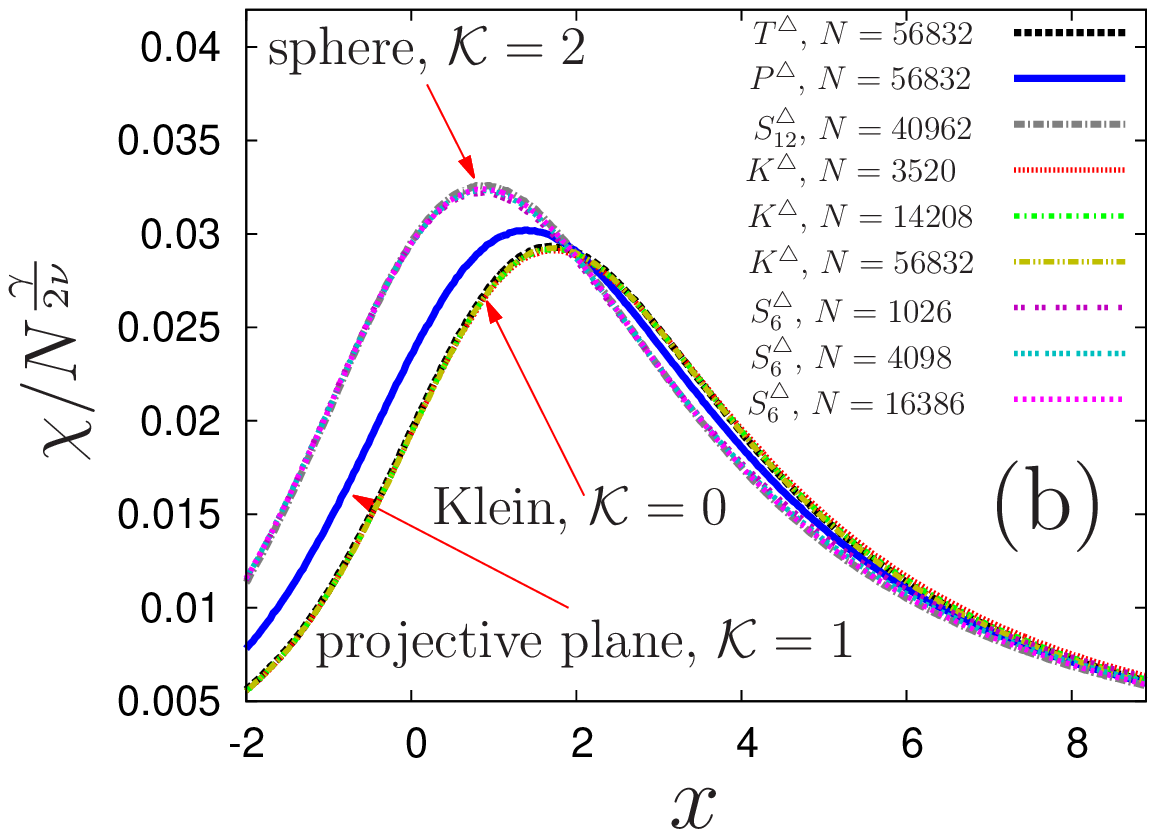}
\includegraphics[width=0.49\textwidth]{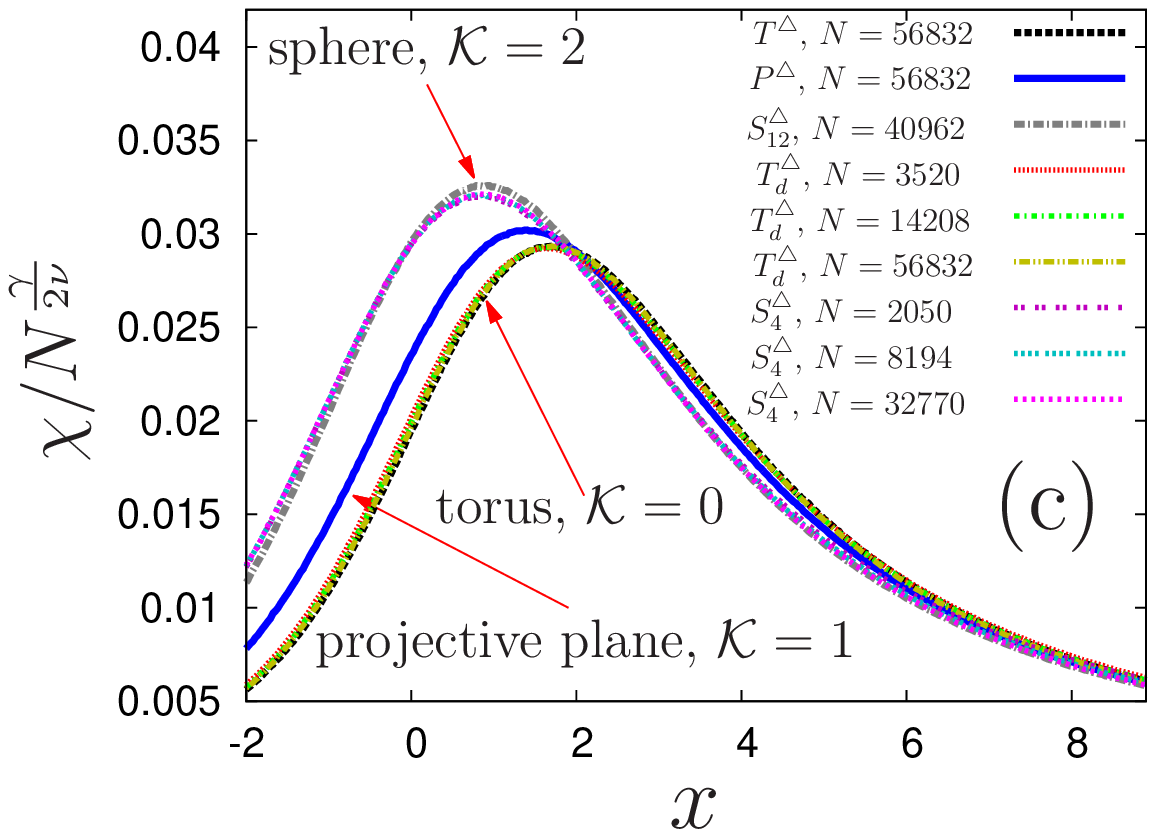}
\includegraphics[width=0.49\textwidth]{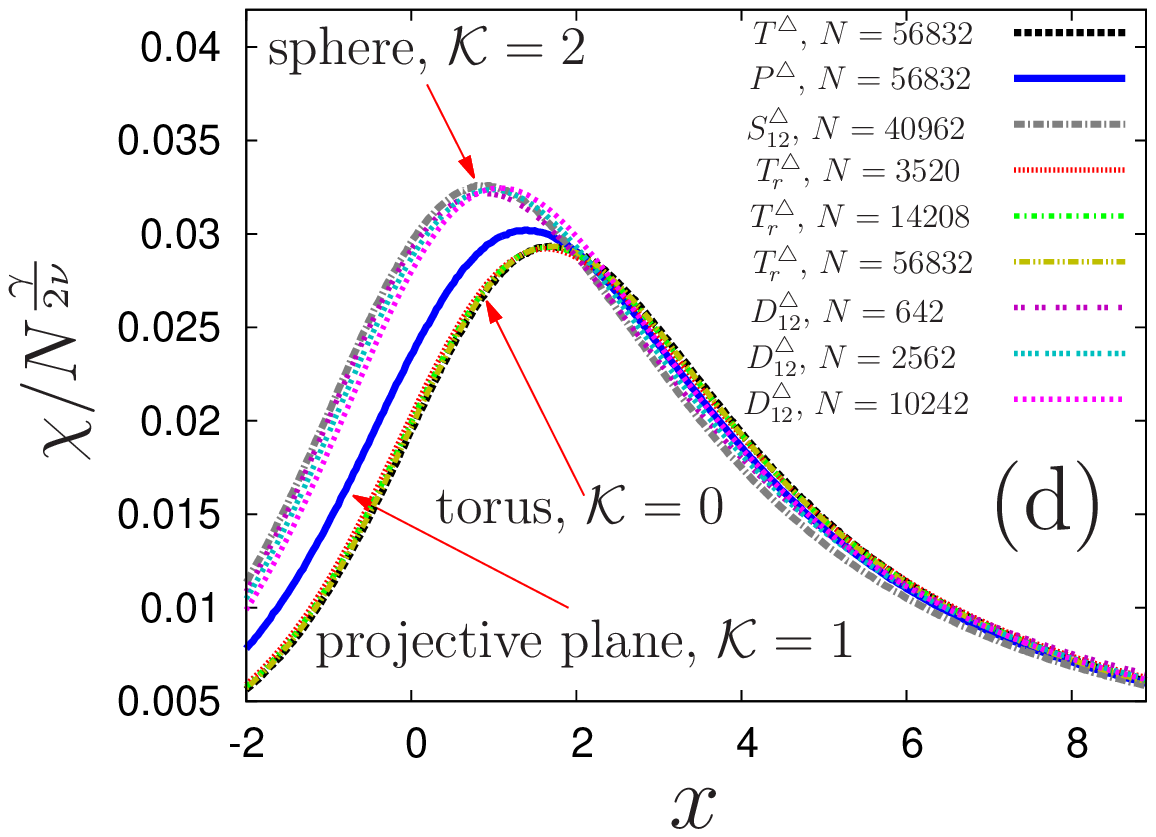}
\caption{
Rescaled magnetic susceptibility $\chi/N^{ \gamma/(2\nu)}$
for triangular lattice decorations
as a function of the scaling variable $x=t( N^{1/2}/\xi_0^+)^{1/\nu}$
for various topologies (the reference curves are $T^{\triangle}$ and $P^{\triangle}$ 
for $N=56832$ and $S^{\triangle}_{12}$ for $N=40962$, which are reproduced in all panels):
(a)  torus ($T^{\triangle}$),  real projective plane ($P^{\triangle}$),
      and regular triangulation
     of a sphere with  $n_5=12$ defects ($S^{\triangle}_{12}$);
(b)  reference curves, curves for the  Klein bottle ($K^{\triangle}$), 
     and curves for the regular triangulation
     of a sphere with $n_4=6$ defects ($S^{\triangle}_6$);
(c)  reference curves, curves for the torus with regularly \textit{d}istributed defects denoted by subscript $d$ ($T^{\triangle}_d$),
     and  curves for the regular triangulation
     of a sphere with $n_3=4$ defects ($S^{\triangle}_4$);
(d)  reference curves,  curves for the torus with \textit{r}andomly distributed defects denoted by subscript $r$ ($T^{\triangle}_r$),
     and  curves for the random  triangulation
     of a sphere with the number of particles as for $n_5=12$ defects ($D^{\triangle}_{12}$).
}
\label{fig:chi_com}
\end{figure*}

In Fig.~\ref{fig:chi_com} we plot the rescaled magnetic susceptibility in terms of the
magnetic susceptibility scaling function ${\mathcal X}=\chi/N^{\gamma/(2 \nu)}$
as  function of the temperature scaling variable $x$ for the triangular ($\triangle$)
lattice, forming surfaces with different topologies  and  for several  numbers of spins $N$. 
As  master curves we select the regular toroidal geometry 
($T^{\triangle}$) with $N=56832$ spins (dashed line), the
real projective plane geometry 
($P^{\triangle}$) with $N=56832$ spins (solid line),
 and the regular sphere 
triangulation ($S^{\triangle}_{12}$) with $N=40962$ spins (dash-dotted line).
As a reference, these master curves are included in all panels.

For completeness, in Fig.~\ref{fig:chi_com}(a) we reproduce our results from Ref.~\cite{VMD}, which  allows us to compare the regular toroidal geometry
($T^{\triangle}$), the projective plane geometry ($P^{\triangle}$),
and the regular triangulation of the sphere with icosahedron symmetry ($S^{\triangle}_{12}$).
In Fig.~\ref{fig:chi_com}(b) we focus on the comparison of  the Klein bottle ($K^{\triangle}$)
with  the regular triangulation 
of the sphere with octahedron symmetry ($S^{\triangle}_{6}$).
In Fig.~\ref{fig:chi_com}(c),   the toroidal geometry
with equidistant defects ($T_d$)  is  compared with the regular triangulation
of a sphere with tetrahedron symmetry ($S^{\triangle}_{4}$).
Finally, in Fig.~\ref{fig:chi_com}(d) we compare  the torus  with  randomly distributed defects ($T_r$) and  the
 sphere ($D^{\triangle}_{12}$) with random triangulation.
We find, that the magnetic susceptibility always falls onto  master curves
defined by the Euler characteristic.
 
\begin{figure*}[h]
\includegraphics[width=0.49\textwidth]{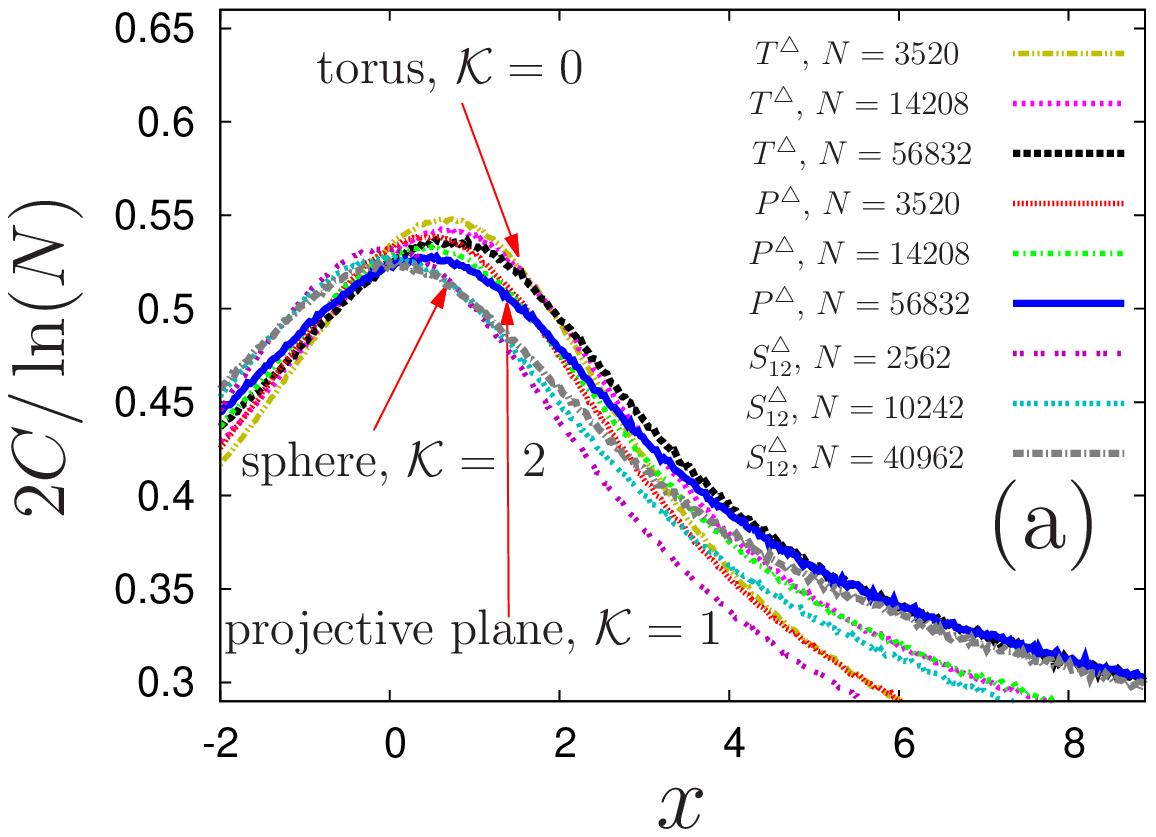}
\includegraphics[width=0.49\textwidth]{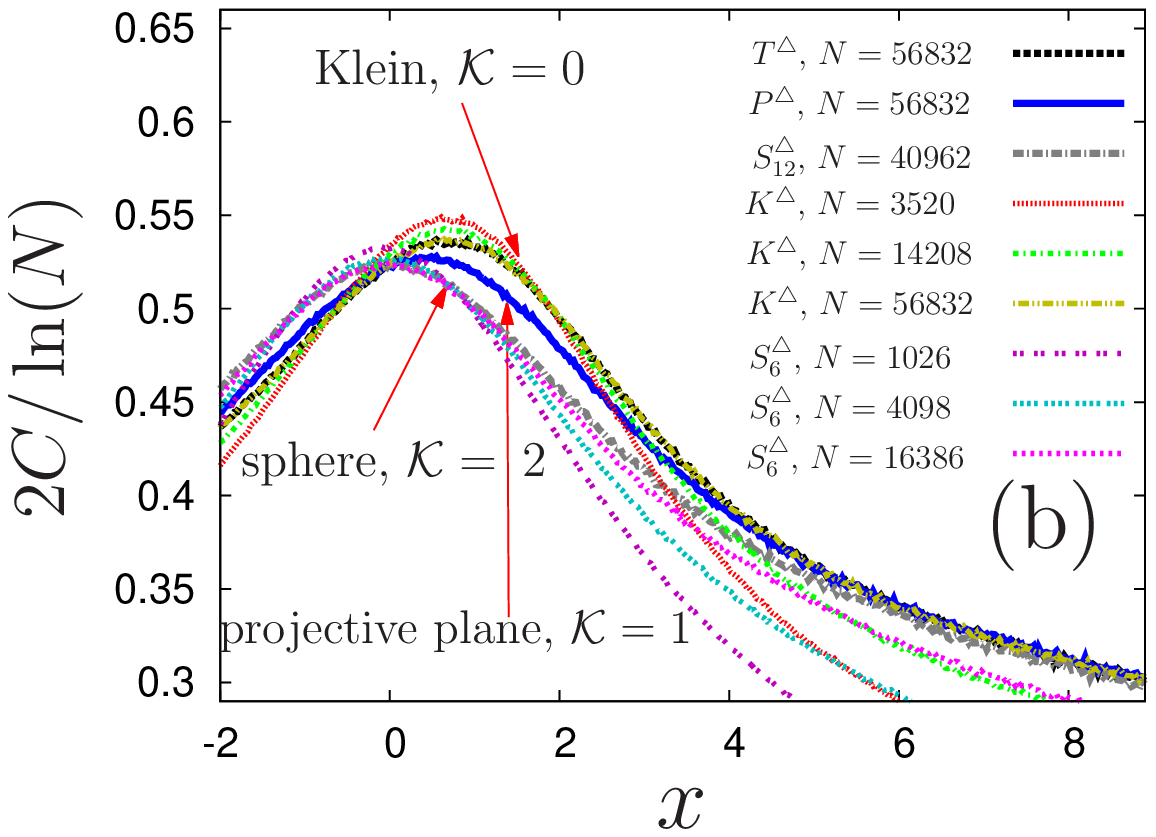}
\includegraphics[width=0.49\textwidth]{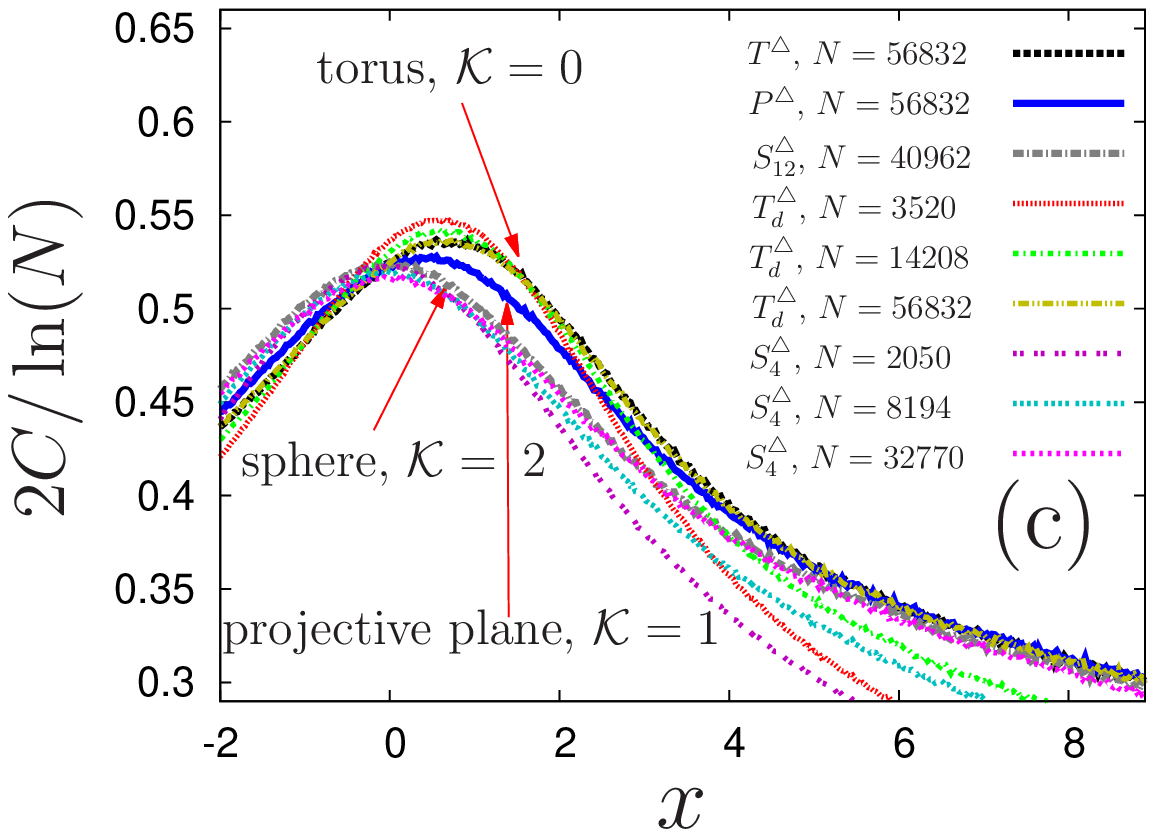}
\includegraphics[width=0.49\textwidth]{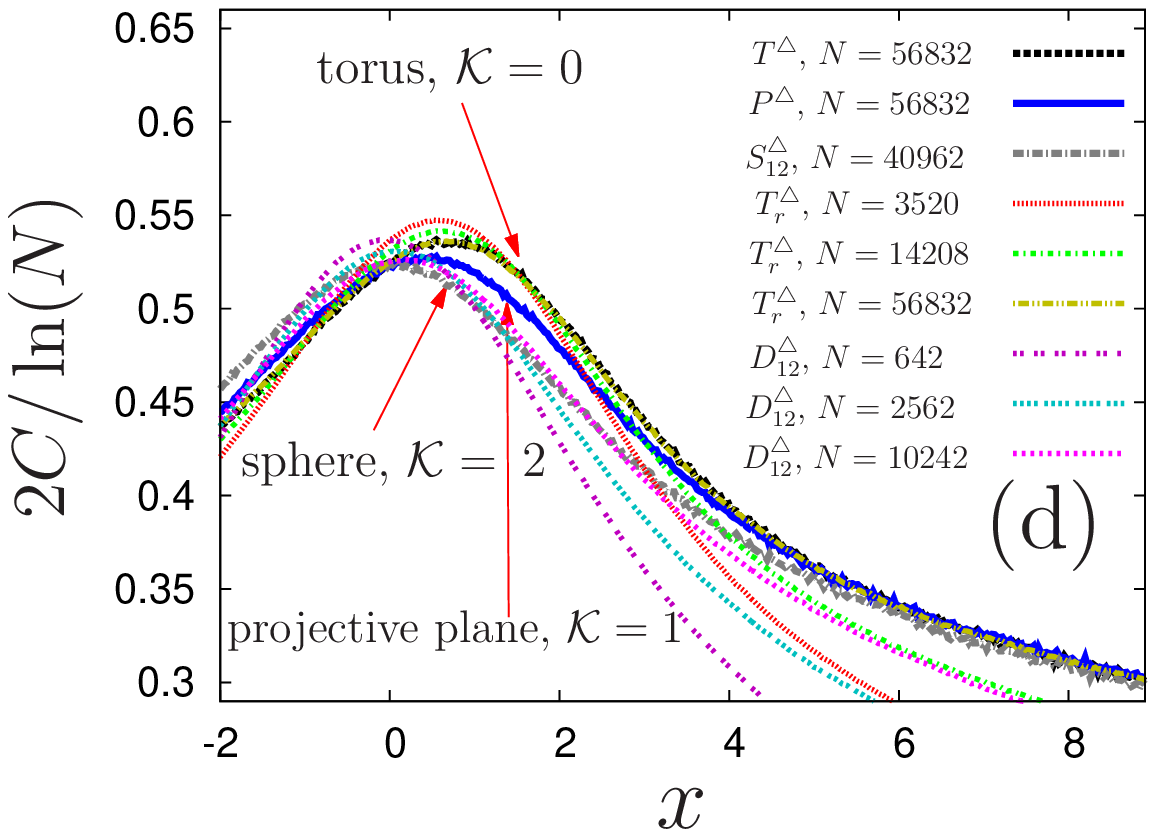}
\caption{
Rescaled specific heat $2 C/\ln(N)$
for triangular lattice decorations
as  function of the scaling variable $x=t( N^{1/2}/\xi_0^+)^{1/\nu}$
for various types of topologies ( the reference curves are $T^{\triangle}$ and $P^{\triangle}$
for $N=56832$ and $S^{\triangle}_{12}$ for $N=40962$):
(a)  torus ($T^{\triangle}$),  real projective plane ($P^{\triangle}$),
      and regular triangulation
     of a sphere with  $n_5=12$ defects ($S^{\triangle}_{12}$);
(b)  reference curves, Klein bottle ($K^{\triangle}$), 
     and regular triangulation
     of a sphere with $n_4=6$ defects ($S^{\triangle}_6$);
(c)  reference curves, torus with regularly \textit{d}istributed defects ($T^{\triangle}_d$),
     and regular triangulation
     of a sphere with $n_3=4$ defects ($S^{\triangle}_4$);
(d)  reference curves, torus with \textit{r}andomly distributed defects ($T^{\triangle}_r$),
     and random  triangulation
     of a sphere with the number of particles as for $n_5=12$ defects ($D^{\triangle}_{12}$).
}
\label{fig:c_com}
\end{figure*}

 In Figs.~\ref{fig:c_com}(a), (b), (c), and (d) we show  the rescaled 
specific heat $2C/\ln(N)$  as a function of the scaling variable $x$
for  a triangular lattice and for the same geometries  
as in Figs.~\ref{fig:chi_com}(a), (b), (c), and (d), respectively. 
We observe that, unlike the magnetic susceptibility, the specific heat does not ideally fall into a set of  master curves.
This could be  due to large finite size corrections to the logarithmic scaling  and  due 
to the contribution of the regular part 
$C(\beta,N) \simeq C_{\mathrm{reg}}+{\mathcal C}(x) \ln(N)/2$, which defines the scaling function  ${\mathcal C}(x)$  (see Appendix~\ref{appendix:B} for additional details). 
 Nonetheless one can clearly
identify three branches of curves which correspond to systems
with different values  of the Euler characteristic $\mathcal K=0,1,2$.
But the order  of the curves is opposite to that for the magnetic 
susceptibility: upon increasing the Euler characteristic the function  $2C/\ln(N)$ shifts to  lower values and its maximum moves to  smaller values
of the scaling variable $x$;  
 the curves from top to bottom in Fig.~\ref{fig:c_com} correspond to the  torus and the Klein bottle,  to the projective plane, and to the  sphere. 
 For the curves in Fig.~\ref{fig:chi_com} the opposite holds.

\begin{figure*}[h]
\includegraphics[width=0.49\textwidth]{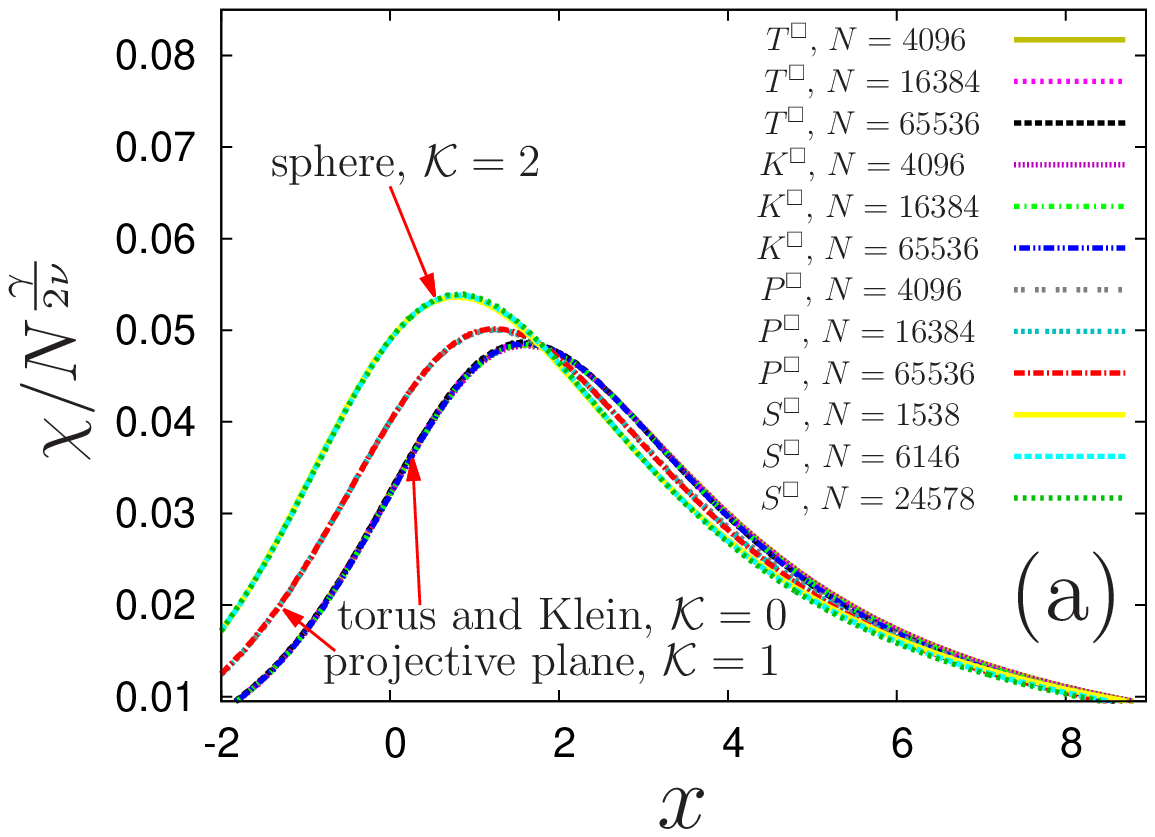}
\includegraphics[width=0.49\textwidth]{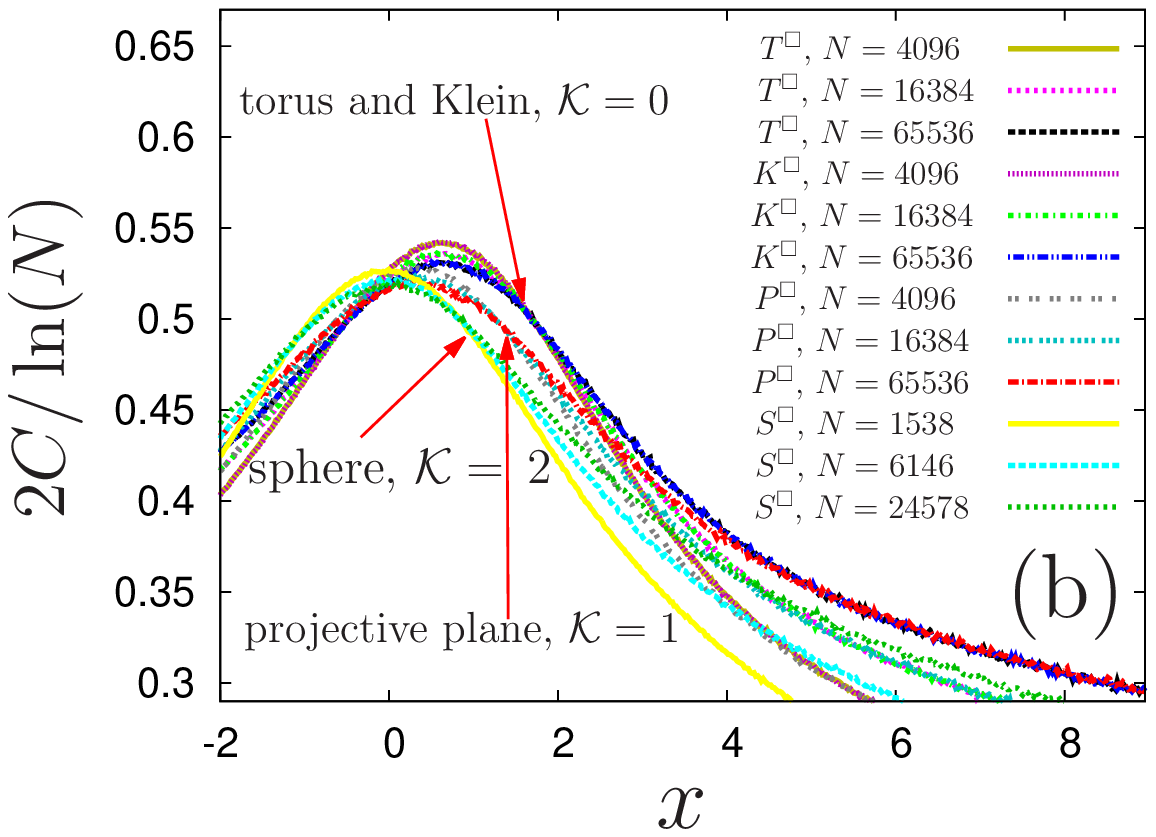}
\caption{
Simulation results for square lattice decorations with various topologies:
torus ($T^{\square}$), Klein bottle ($K^{\square}$), 
real projective plane ($P^{\square}$),
and deformed cube surface ($S^{\square}$).
(a)  Rescaled magnetic susceptibility $\chi/N^{ \gamma/(2\nu)}$ and (b)  rescaled specific heat $2 C/\ln(N)$
as  function of the scaling variable $x=t( N^{1/2}/\xi_0^+)^{1/\nu}$.
}
\label{fig:sq}
\end{figure*}

Again for completeness, in  Fig.~\ref{fig:sq}(a) we reproduce results from Ref.~\cite{VMD} for the magnetic susceptibility scaling function for the square lattice decoration 
with various topologies and system sizes. 
As for the triangular lattice, we observe a
perfect data collapse for the surfaces with the same value of the 
Euler characteristic $\mathcal K$.

The specific heat for the square lattice and for various topologies is
 shown in Fig.~\ref{fig:sq}(b). It  behaves in the same way  as 
the specific heat for the triangular decorations.

\begin{figure*}[h]
\includegraphics[width=0.49\textwidth]{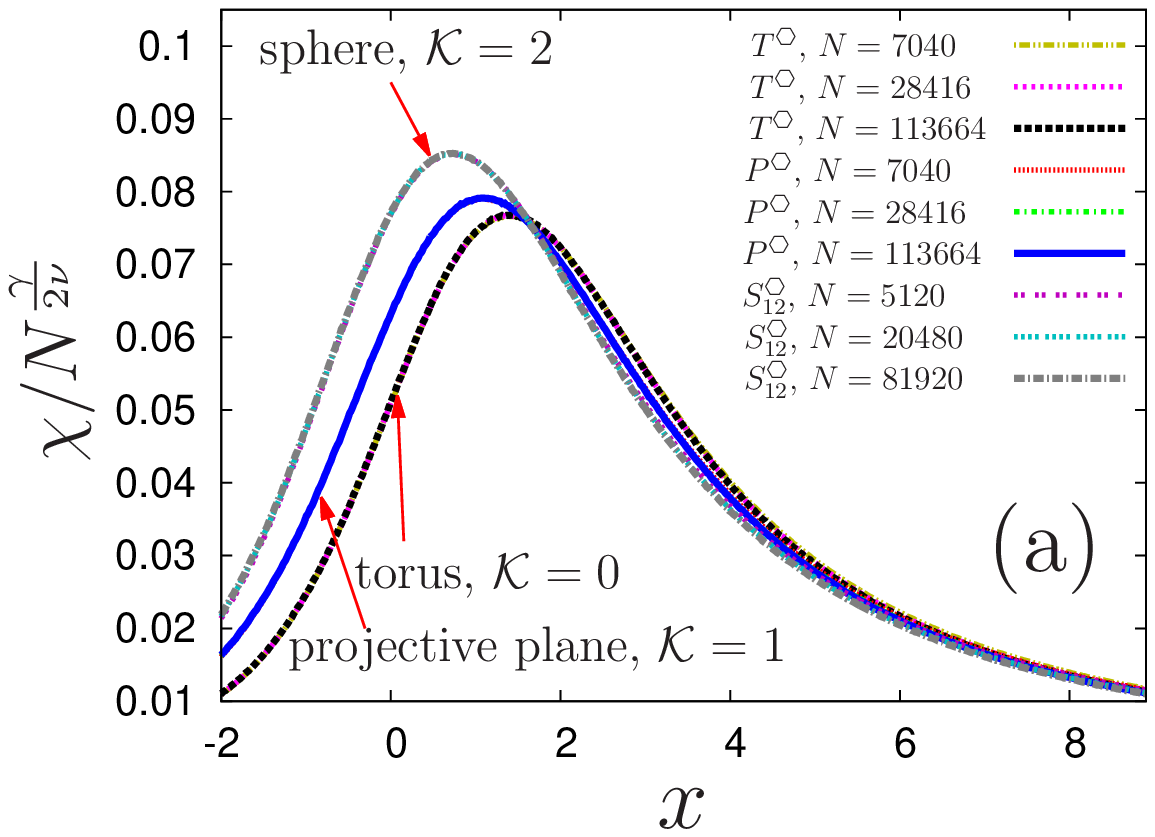}
\includegraphics[width=0.49\textwidth]{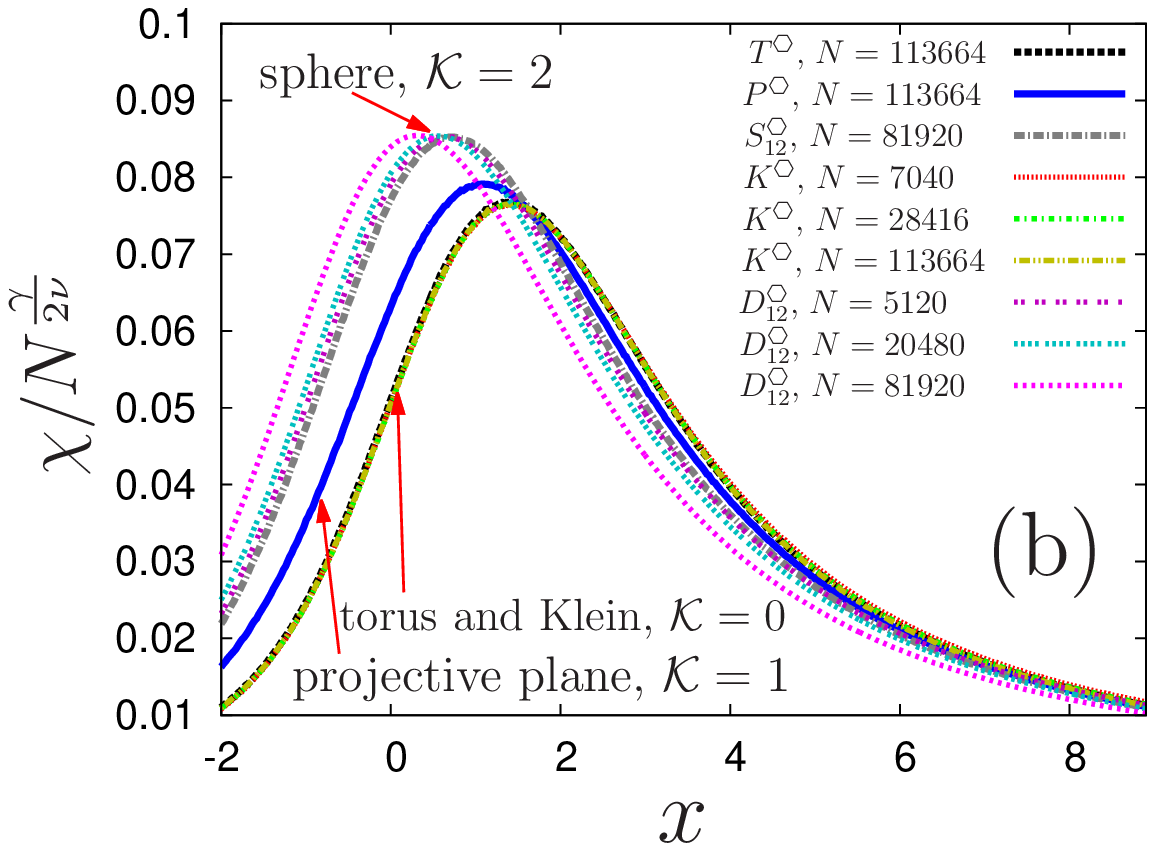}
\caption{
Rescaled magnetic susceptibility $\chi/N^{ \gamma/(2\nu)}$
for hexagonal lattice decorations
as   function of the scaling variable $x=t( N^{1/2}/\xi_0^+)^{1/\nu}$ and
for various  topologies. The reference curves are $T^{\hexagon}$ and $P^{\hexagon}$
for $N=113664$ and $S^{\hexagon}_{12}$ for $N=81920$.
(a)  Torus ($T^{\hexagon}$),  real projective plane ($P^{\hexagon}$),
      and  sphere  ($S^{\hexagon}_{12}$, dual to $S^{\triangle}_{12}$);
(b)  reference curves, Klein bottle ($K^{\hexagon}$),
 and random  representation of a sphere  ($D^{\hexagon}_{12}$ dual to $D^{\triangle}_{12}$).}
\label{fig:chi_hex}
\end{figure*}

In Fig.~\ref{fig:chi_hex} we present simulation results for  the  magnetic susceptibility in the  case of a lattice with hexagonal decoration.
The simulation results for the regular torus ($T^{\hexagon})$,  the real projective plane ($P^{\hexagon}$), and the sphere $S^{\hexagon}_{12}$
(dual to $S^{\triangle}_{12}$) are presented in Fig.~\ref{fig:chi_hex}(a).
Finally, in Fig.~\ref{fig:chi_hex}(b) we compare the master curves from Fig.~\ref{fig:chi_hex}(a)
with the results for the  Klein bottle $K^{\hexagon}$  and
 the dynamic configuration  $D^{\hexagon}_{12}$ (which is dual to the
dynamic triangulations $D^{\triangle}_{12}$).

In sum, we find the same feature  as for the other decorations: the 
curves split into three branches which correspond to  three different  values 
of the  topological characteristic, \textit{i.e.}, $\mathcal K=0,1,2$.
The results for the torus and for the  Klein bottle  belong to the same branch.

\section{Conclusions}

We have numerically investigated the dependence
of thermodynamic functions of the two-dimensional Ising model on the
topology of its surfaces. We have studied lattices with triangular, square, and hexagonal  facets.
These lattices have the topology of a torus (Euler characteristic $\mathcal K=0$), a
Klein bottle ($\mathcal K=0$),  a real projective plane ($\mathcal K=1$),
and  a sphere ($\mathcal K=2$), respectively. We find, that the magnetic susceptibility scaling function
splits into distinct  branches of curves in accordance with the topology classes 
of the surfaces. For each type of the lattice tiling and within the same topological class (\textit{i.e.},  the same value 
of the Euler characteristic $\mathcal K$) the curves exhibit excellent data collapse, while 
for different topological classes the  curves differ  from each other. There are differences between the shapes of the scaling functions for the same topology but different
lattice tiling. Importantly, we have found that also the scaling function of the specific heat depends on the topological class of the surface. 
Although  in this case the data collapse is  much worse, we can firmly state that there is a similar splitting into branches
associated with the value of $\mathcal{K}$.

A natural but challenging extension of the present research would be
off-lattice simulations  using, for example, molecular dynamics methods.
This would allow one to address the issue of universality for systems
confined to surfaces with different topologies. It is unclear  to which extent  the
universal properties of the critical demixing transition of binary 
liquid mixtures are shared by the two-dimensional Ising model if the
surfaces of these  systems are nonplanar.

\begin{appendix}
\section{Aspect ratio dependence and finite size scaling for the magnetic succeptibility}
\label{appendix:aspect}

The general form of the aspect ratio dependence of  the percolation
function $\pi_{h}(\rho)$ (\textit{i.e.}, the probability to have a percolating cluster
in the \textit{h}orizontal direction) has been presented by Langlands \textit{et al.}~\cite{LPPS}.
We note that the percolation temperature of so-called physical clusters in the 2D Ising model with vanishing bulk ordering field coincides
with the critical temperature $T_c$. Therefore   the percolation probability behaves in a way very similar to the
spontaneous magnetization~\cite{CNPR}. 
 For each type of the lattice, there exists a special, so-called ``reference'' value $r_0$ of the aspect ratio  $\rho=L_{1}/L_{2}$, 
which captures the geometrical size of the lattice. Leglands showed  that if one measures the aspect ratio $\rho$ in units of $r_0$,
\textit{i.e.}, if one writes $\rho=r r_{0}$ where $r\equiv (L_{1}/L_{2})/r_0$, 
the crossing probability as a function of $\rho$ may be written as 
\begin{equation}
 \label{eq:Leglands}
 \pi_{h}(r r_{0})=\eta_{h}(r)
\end{equation}
 where
$\eta_{h}(r)=\eta_{h}(1/r)$ is  a universal function 
(which is self-dual with respect to its argument $r \to 1/r$).
The values of this reference aspect ratio 
for triangular, square, and hexagonal lattices are
$r_{0}=\sqrt{3}/2,1,$ and $\sqrt{3}$, respectively. 
We recall, that the system sizes $L_1$ and $L_2$
are measured in numbers of spins in the rows and  the columns, respectively.
 The result in Eq.~(\ref{eq:Leglands}) implies, that for the comparison of the scaling functions for different types of lattices, one should choose
$L_1$ and $L_2$ such that the geometrical sizes  of these lattices are equal. 
In view of Eq.~(\ref{eq:Leglands}) we expect that Eq.~(\ref{eq:chi_scal_sq}) may be written for various lattice types 
as
\be
\label{eq:chi_scal}
\chi(\beta,r r_{0} L,L)=N^{\frac{\gamma}{2 \nu}}
 {\mathcal X}(\tau, r),
\ee 
where the scaling function ${\mathcal X}$
is dual with respect to its second argument: 
${\mathcal X}(\tau,r)={\mathcal X}(\tau,1/r)$.   
%The dual aspect ratio $\rho'$ which corresponds to the value $\rho=r r_{0}$ may be expressed as $\rho'=r_{0}/r=r_{0}^{2}/\rho$.
  
\begin{figure*}[h]
\centerline{
\mbox{\includegraphics[width=0.49\textwidth]{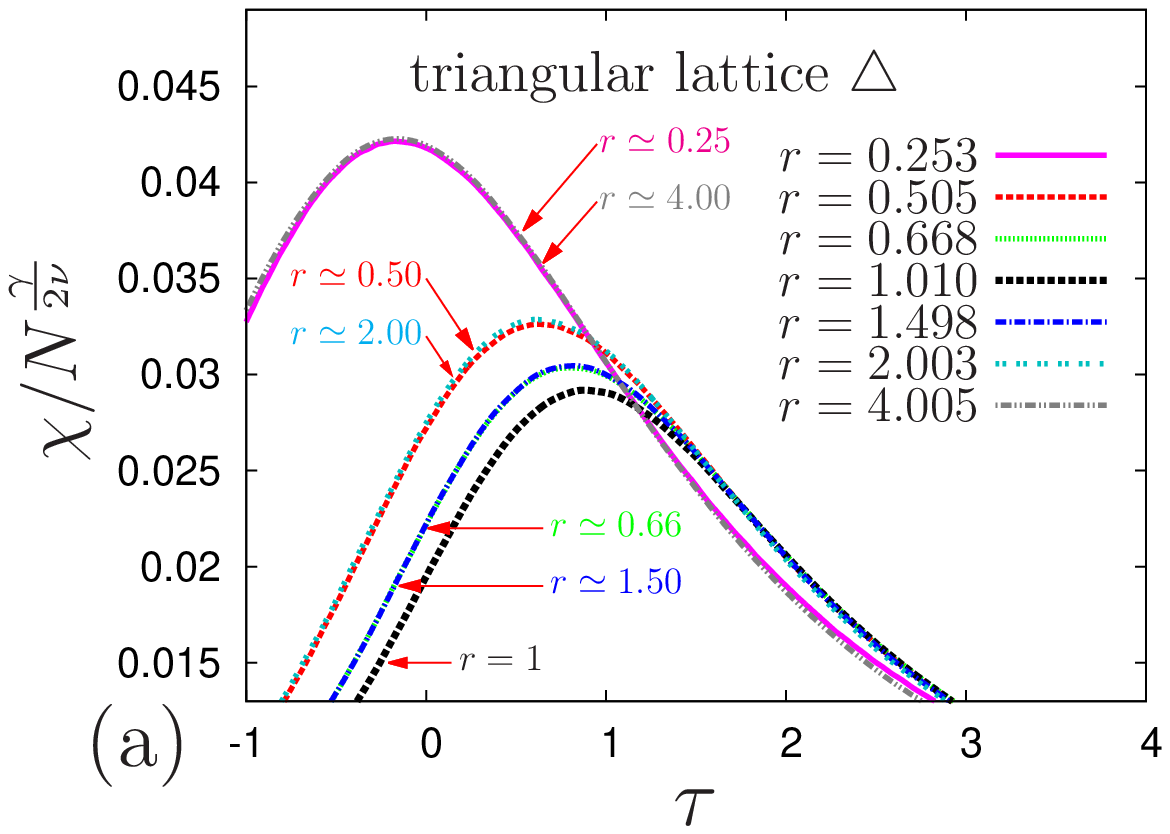}\hspace{-0.02\textwidth}
\includegraphics[width=0.49\textwidth]{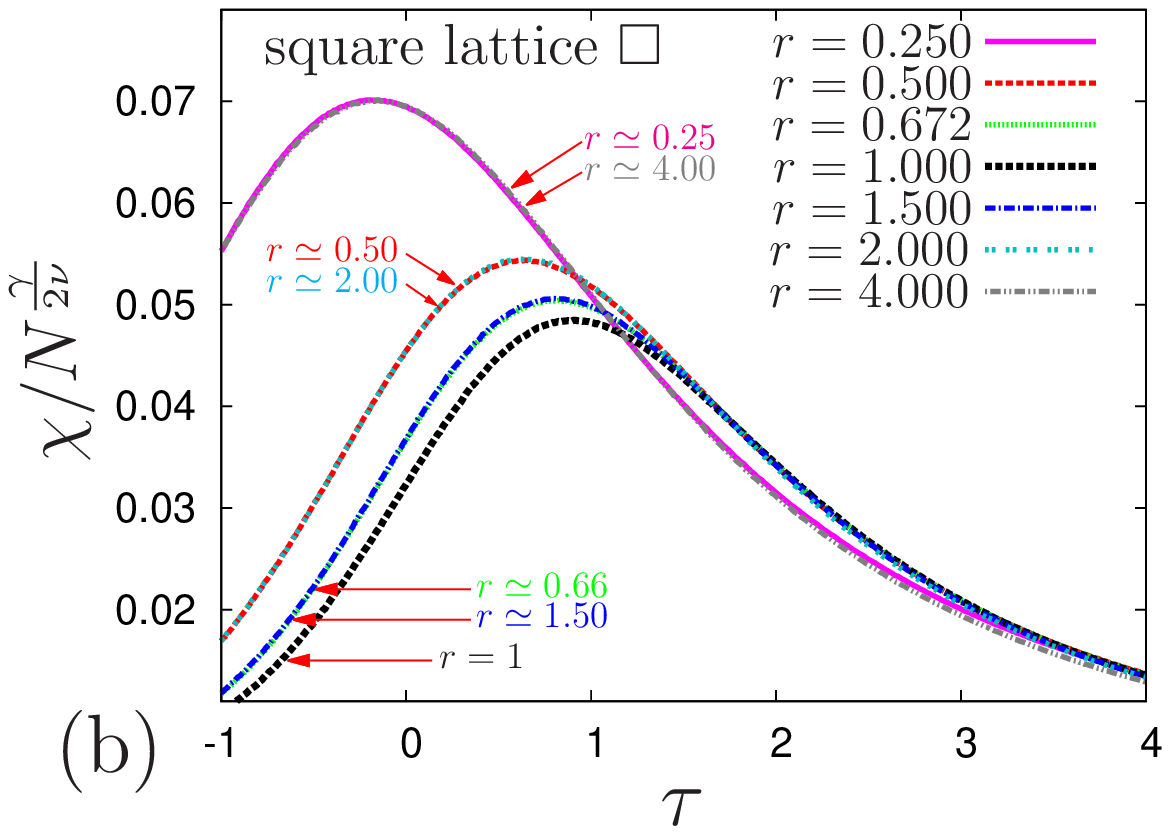}}}
\centerline{
\mbox{
\includegraphics[width=0.49\textwidth]{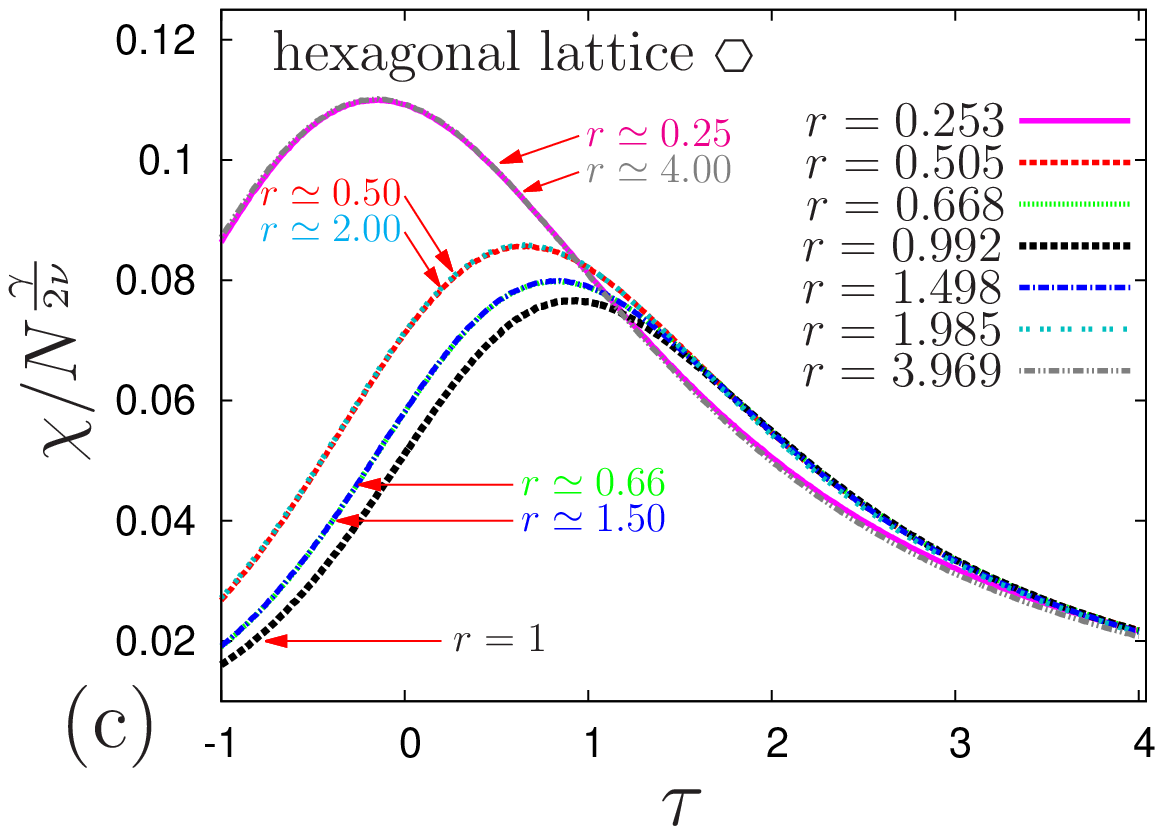}
\includegraphics[width=0.49\textwidth]{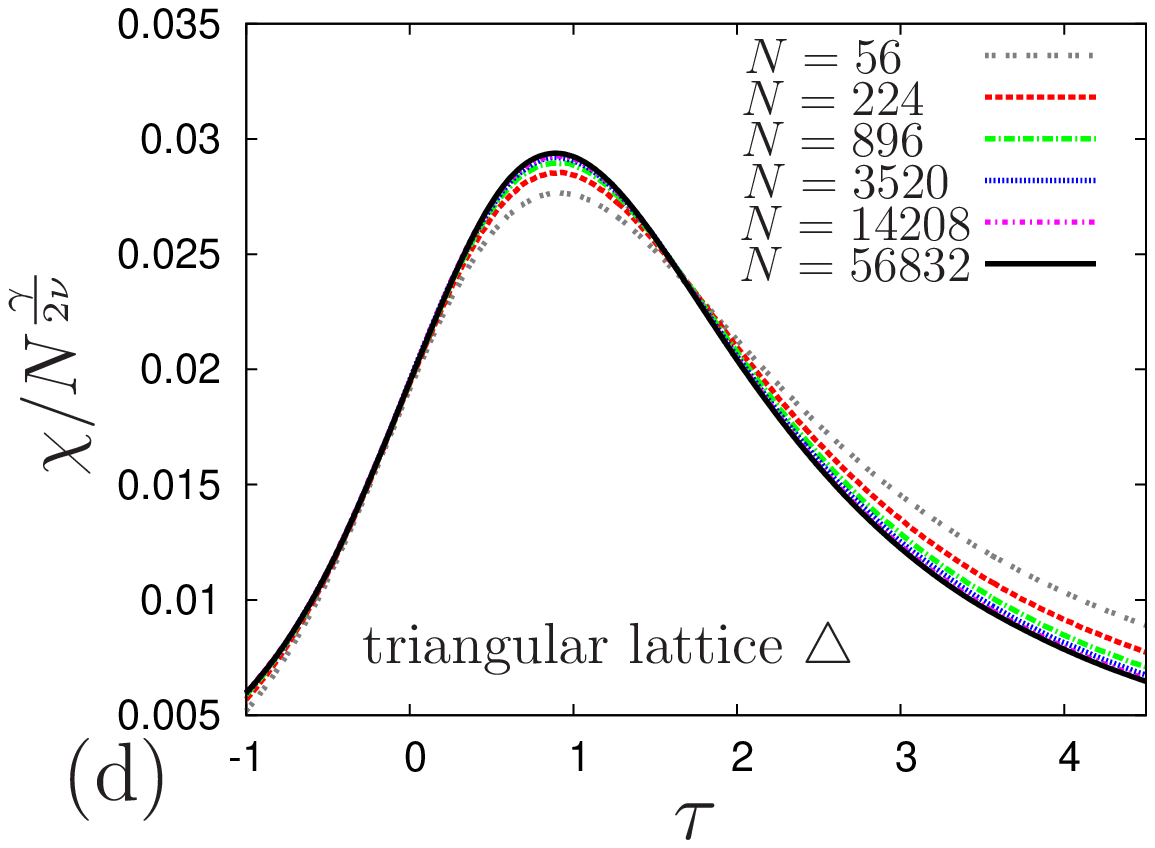}}}
\caption{
Rescaled magnetic susceptibility $\chi/N^{ \gamma/(2\nu)}$
as  function of the temperature scaling variable $\tau=t N^{1/(2 \nu)}$. (The temperature scaling variable $x$ includes the correlation
   length amplitude $\xi_0^+$, the variable $\tau$ does not include it: $x=\tau(\xi_0^+)^{-1/\nu}$.)  The data correspond to
 the toroidal topology with various values of the reduced aspect  ratio $r=\rho/r_0$:
(a)  on the triangular lattice;
(b)  on the square lattice;
(c)  on the hexagonal lattice.
(d) Rescaled magnetic susceptibility $\chi/N^{ \gamma/(2\nu)}$
as  function of the scaling variable $\tau=t N^{1/(2 \nu)}$
on the triangular lattice ({\it i.e.}, the toroidal topology)
 for various numbers of spins: $N=56,224,896,3520,14208$, and $56832$; 
the aspect ratio is  $\rho \simeq r_{0}=\frac{\sqrt{3}}{2}$.  For further information see the main text in Appendix A.
}
\label{fig:chi_asp}
\end{figure*}

We have checked Eq.~(\ref{eq:chi_scal}) numerically and present the corresponding results
 in Fig.~\ref{fig:chi_asp}. In this figure we plot 
 the rescaled magnetic susceptibility ${\mathcal X }= \chi/N^{\gamma/(2 \nu)}$
 as function of the temperature scaling variable $\tau=t N^{1/(2 \nu)}$
for  triangular [Figs.~\ref{fig:chi_asp}(a)  and \ref{fig:geom_plane}(a)], 
square [Figs.~\ref{fig:chi_asp}(b)  and \ref{fig:geom_plane}(b)], and hexagonal [Figs.~\ref{fig:chi_asp}(c) and 
\ref{fig:geom_plane}(c)] lattices.
These lattices have a rectangular shape of  width $L_{1}=64$ and periodic 
boundary conditions (i.e., toroidal geometry).
The considered values of the reduced aspect ratio $r=\rho/r_0$ (\textit{i.e.,} $\rho$ in units of $r_0$) 
are $r=0.25,0.5,0.6\bar{6},1,1.5,2,$  and $4$. 
The length of the lattice
$L_{1} \simeq \rho L_{2}= r r_{0} L_{2}$ is chosen  as the integer
closest to the number $r r_{0} L_{2}$. In every panel we plot 
the magnetic susceptibility for  the self-dual ratio $r \simeq 1$
and for a set of dual values $r$ and $r' \simeq 1/r$. (In the general case 
it is impossible to realize the equality $r' = 1/r$ because both $L_{1}$ and $L_{2}$ 
are integer.)  Upon increasing the reduced aspect ratio $r$, the position $\tau_{\mathrm{max}}$ of the maximum of $\chi/N^{\gamma/(2\nu)}$ as a 
 function of $r$ shifts to larger values of $\tau$. This occurs up to the self-dual value $r=1/r=1$. Upon  increasing $r$ further, the position of the maximum shifts back to lower values of $\tau$. 
 We observe that  the values of $\chi/N^{\gamma/(2\nu)}$ for $r$ and $r' \simeq 1/r$ coincide in accordance with  our expectation that ${\mathcal X}(\tau,r)={\mathcal X}(\tau,1/r)$.
Thus, for the triangular and the hexagonal lattice we should use the aspect ratio, which is  a multiple of the reference value 
$ r_{0}=\sqrt{3}/2 \quad  \simeq 0.86602$ and $ r_{0}=\sqrt{3} \quad \simeq 1.73205$, respectively, 
and which corresponds to the square geometry on the square lattice.

In Fig.~\ref{fig:chi_asp}(d)
we plot the rescaled magnetic susceptibility $\chi/N^{ \gamma/2\nu}$
as a function of the scaling variable $\tau=t N^{1/(2 \nu)}$
on the triangular lattice with the selected aspect ratio $\rho \simeq r_{0}=\sqrt{3}/2 $  
for various numbers of spins: $N=56,224,896,3520,14208, \mathrm{and} \quad 56832$.
We observe noticeable finite size corrections to scaling for 
lattices of small sizes ($N<1000$) while for larger system sizes the curves coincide.
  Therefore it is preferable 
to use large lattices ($N >1000$) in order to avoid 
corrections to scaling and  to observe good data collapse.

\section{Finite size scaling for the specific heat}
\label{appendix:B}

\begin{figure*}[h]
\centerline{
\includegraphics[width=0.45\textwidth]{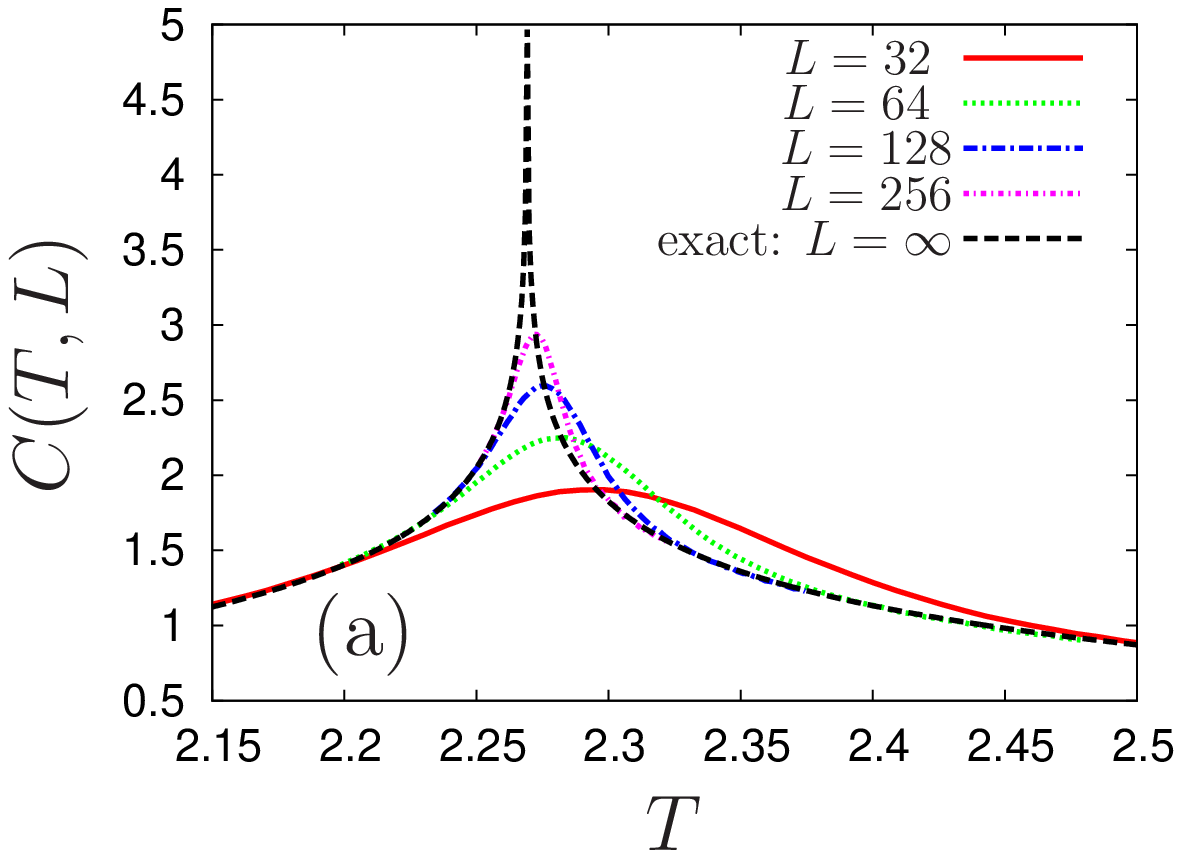}
\includegraphics[width=0.45\textwidth]{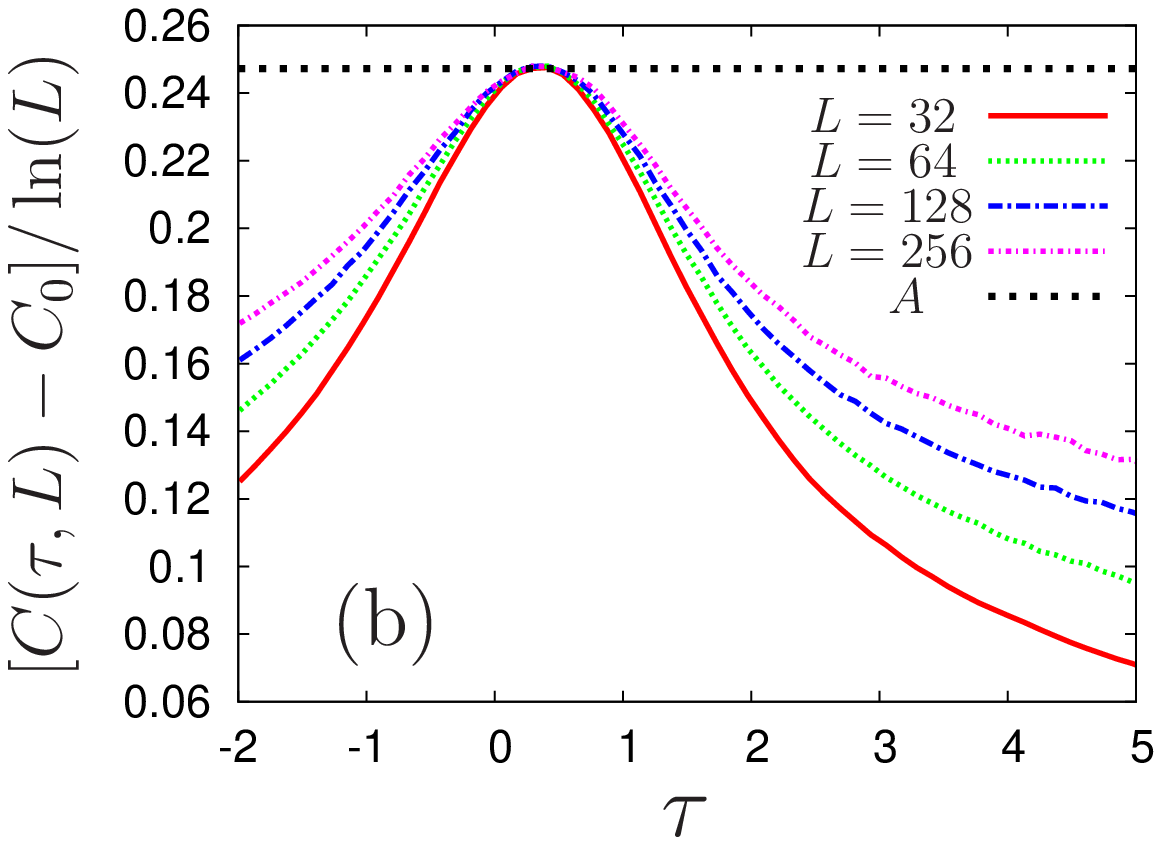}
}
\caption{
(a) Specific heat $C(T,L)$ as  function of the temperature
$T$ for four lattice sizes $L=32,64,128,256$
in comparison with the exact result in Eq.~(\ref{eq:c_exact}).
(b) Rescaled specific heat $(C-C_{0})/\ln(L)$ as  function
of $\tau=t L$ compared  with the maximum value $A$.  For further information see the main text in Appendix B.
}
\label{fig:c_scal}
\end{figure*}

The singular part of the specific heat of the Ising model on the two-dimensional lattice exhibits a logarithmic divergence.
In Fig.~\ref{fig:c_scal}(a) we compare numerical results
for  the specific heat $C$ as  function of temperature $T$ (in units of the coupling constant $J$)
on  square lattices of sizes $L=32,64,128,$ and $256$  for the
 toroidal geometry
with the analytic expression \cite{MW}
\be
\label{eq:c_exact}
C(T\to T_c) \simeq \frac{8}{\pi T_{c}^{2}} \left[ \ln \left|
\frac{\sqrt{2}}{\frac{1}{T}-\frac{1}{T_{c}}} \right| -(1+\pi/4)\right]
\ee

Concerning the maximum of the specific heat for finite $L$ one has
$C_{\mathrm{max}}(L)\simeq A \ln(L) + C_{0} $ where $A=\frac{4}{\pi T_{c}^2}$
and $C_{0}\simeq 0.1879\frac{1}{T_{c}^2}$. In Fig.~\ref{fig:c_scal}(b) we plot the
rescaled specific heat $[C-C_{0}]/\ln(L)$ as  function of the scaling variable
$\tau=t L$ in comparison with $A$.
We find, that after this rescaling all maxima coincide
and correspond to the expected value $A$. However, in Fig.~\ref{fig:c_scal}(b) the
curves differ from each other and it is impossible
to bring them  on top of each other  by rescaling  the axis $\tau$.
The  issue of  how to rescale the specific heat with the logarithm $\ln(L)$ of the system size,
 in order to obtain  data collapse,  remains open. We have decided to scale our  results for the  specific heat as
$C/\ln(L) =2 C/\ln(N)$, where the total number $N$  of spins  for the square lattice is $N=L^2$.
The aspect ratio dependence of the specific heat
of the Ising model on the square lattice of rectangular shape
has been studied in Ref.~\cite{FF}.

\end{appendix}

\newpage

\newpage

\begin{figure*}[h]
\includegraphics[width=0.49\textwidth]{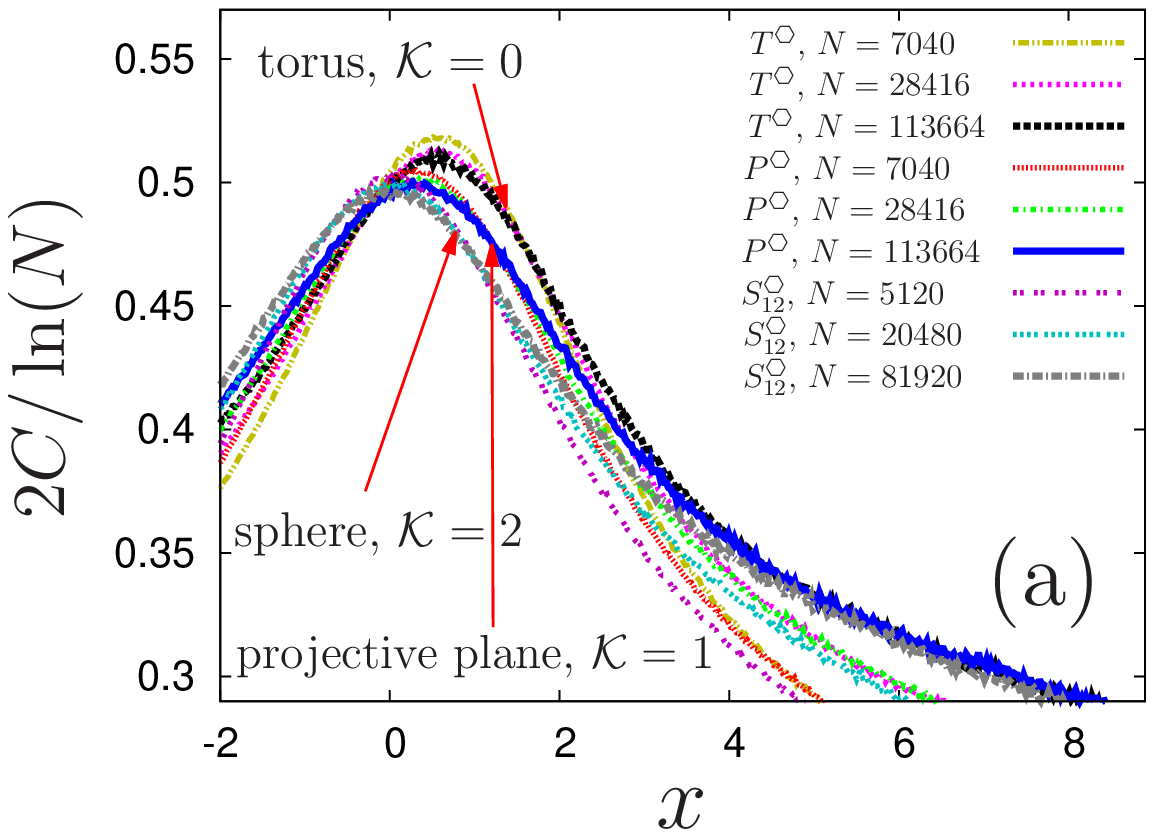}
\includegraphics[width=0.49\textwidth]{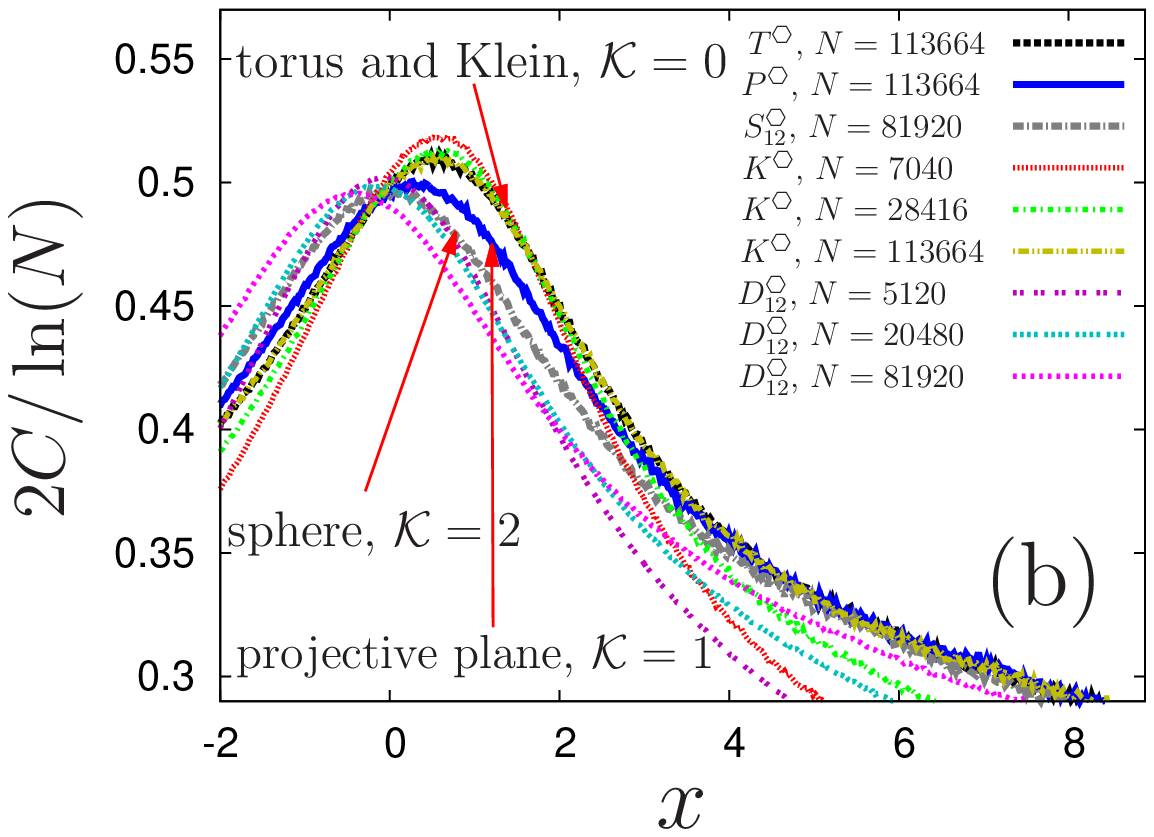}
\caption{
Rescaled specific heat $2 C/\ln(N)$
for hexagonal lattice decorations
as  function of the scaling variable $x=t( N^{1/2}/\xi_0^+)^{1/\nu}$  and
for various topologies. The reference curves are $T^{\hexagon}$ and $P^{\hexagon}$
for $N=113664$ and $S^{\hexagon}_{12}$ for $N=81920$.
(a)  Torus ($T^{\hexagon}$),  real projective plane ($P^{\hexagon}$),
      and  sphere  ($S^{\hexagon}_{12}$, dual to $S^{\triangle}_{12}$);
(b)  reference curves, Klein bottle ($K^{\hexagon}$),
 and random  representation 
     of a sphere  ($D^{\hexagon}_{12}$ dual to $D^{\triangle}_{12}$).}
\label{fig:c_hex}
\end{figure*}

\end{document}